\documentclass[aps,prd,twocolumn,showpacs,floats,floatfix,nofootinbib]{revtex4}
\usepackage{graphicx}
\usepackage{dcolumn}
\usepackage{bm}
\usepackage{graphics}
\usepackage{color}
\usepackage{amssymb}
\usepackage{amsmath}
\usepackage{natbib}
\usepackage{amsmath}
\usepackage{url}
\newcommand{\ba}{\begin{eqnarray}}
\newcommand{\ea}{\end{eqnarray}}
\def\spose#1{\hbox to 0pt{#1\hss}}

\def\lta{\mathrel{\spose{\lower 3pt\hbox{$\mathchar"218$}}
     \raise 2.0pt\hbox{$\mathchar"13C$}}}
\def\gta{\mathrel{\spose{\lower 3pt\hbox{$\mathchar"218$}}
     \raise 2.0pt\hbox{$\mathchar"13E$}}}
\newcommand{\be}{\begin{equation}}
\newcommand{\ee}{\end{equation}}
\newcommand{\bea}{\begin{eqnarray}}
\newcommand{\eea}{\end{eqnarray}}
\newcommand{\ex}{\mbox{e}}
\newcommand{\dd}{\mbox{d}}

\def\setR{\mathbb{R}}
\def\setC{\mathbb{C}}

\newcommand{\ie}{\textsl{i.e., }}

\newcommand{\eg}{\textsl{e.g., }}

\newcommand{\Hu}{{\cal H}}
\newcommand{\Ka}{{\cal K}}

\newcommand{\cs}{{c_{_\mathrm{S}}^2}}
\newcommand{\GN}{G_{_\mathrm{N}}}
\newcommand{\mP}{m_{_\mathrm{Pl}}}
\newcommand{\ns}{n_{_\mathrm{S}}}
\newcommand{\lP}{\ell_{_\mathrm{Pl}}}

\begin{document}
%\title{Constraining classical bouncing models}
\title{A classical bounce: constraints and consequences}

\author{F. T. Falciano}
\email{ftovar@cbpf.br} 
\affiliation{ICRA - Centro Brasileiro de Pesquisas F\'{\i}sicas -- CBPF, rua Xavier Sigaud, 150, Urca, CEP22290-180, Rio de Janeiro, Brazil}
\affiliation{${\cal G}\setR\varepsilon\setC{\cal O}$ -- Institut d'Astrophysique de Paris, UMR7095 CNRS, Universit\'e Pierre \& Marie Curie, 98 bis boulevard Arago, 75014 Paris, France}

\author{Marc Lilley}
\email{lilley@iap.fr} 
\affiliation{${\cal G}\setR\varepsilon\setC{\cal O}$ -- Institut d'Astrophysique de Paris, UMR7095 CNRS, Universit\'e Pierre \& Marie Curie, 98 bis boulevard Arago, 75014 Paris, France}

\author{Patrick Peter}
\email{peter@iap.fr}
\affiliation{${\cal G}\setR\varepsilon\setC{\cal O}$ -- Institut d'Astrophysique de Paris, UMR7095 CNRS, Universit\'e Pierre \& Marie Curie, 98 bis boulevard Arago, 75014 Paris, France}

\begin{abstract}
We perform a detailed investigation of the simplest possible cosmological model in which a bounce can occur, namely that where the dynamics is led by a simple massive scalar field in a general self-interacting potential and a background spacetime with positively curved spatial sections. By means of a phase space analysis, we give the conditions under which an initially contracting phase can be followed by a bounce and an inflationary phase lasting long enough (\ie at least 60--70 e-folds) to suppress spatial curvature in today's observable universe.  We find that, quite generically, this realization requires some amount of fine-tuning of the initial conditions.  We study the effect of this background evolution on scalar perturbations by propagating an initial power-law power spectrum through the contracting phase, the bounce and the inflationary phase. We find that it is drastically modified, both spectrally ($k-$mode mixing) and in amplitude.  It also acquires, at leading order, an oscillatory component, which, once evolved through the radiation and matter dominated eras, happens to be compatible with observational data.
\end{abstract}
\maketitle
\section{Introduction}
The inflationary paradigm is now part of the standard cosmological model (see Ref.~\cite{inf25} for a recent review with updated references). So much so that the very idea of challenging its main hypothesis might appear hopeless in view of its impressive success in explaining the otherwise mysterious observations that the Universe is flat and seemingly free of remnants such as monopoles and in providing a mechanism that exponentially damps any initial anisotropy~\cite{Turner83b}, as well as considerably alleviates the homogeneity problem, \ie instead of demanding homogeneity over an almost infinite spacelike hypersurface, it reduces to finding a Planck size region over which the inflaton field is homogeneous.  In the usual (chaotic) scenario~\cite{Linde:1981mu,Linde83}, this is almost certain to occur, taking into account the infinite amount of possible initial conditions.  It then suffices to argue that we happen to live in what has become of this initial region (\ie applying some sort of anthropic principle). Moreover, specific examples have shown that an initially inhomogeneous universe, of the Tolman-Bondi type for instance~\cite{CS92}, also shows a trend to homogeneize under the action of inflation. Therefore, and even though these arguments may not be entirely convincing in the most general case, it is plausible to argue, somehow, that inflation provides a dynamical means to drive any ``weird'' universe into one satisfying the cosmological principle. Finally, inflation also predicts the spectrum of primordial density fluctuations thanks to which large-scale structures formed; this predicted spectrum can easily be made to fit all the known data.

Why, then, would one insist in finding an alternative?

First of all, having a serious competitor usually boosts the understanding and the predictability of the defender. Although it is certainly of interest to provide new tests for the inflationary paradigm to confront, it is of crucial importance to provide {\it a priori} alternatives were the inflationary predictions unable to match future observations. In order to do so, one naturally turns to high energy extensions of the standard theory. Nowadays, this mostly means looking at the cosmological consequences of models that can be implemented in string theory (see, \eg \cite{Polchinski98}) in a satisfactory way.

Moreover, inflation itself is not free from difficulties (see for instance the discussion by R.~Brandenberger in Ref.~\cite{inf25}).  Aside from assuming an otherwise never observed scalar field to lead the dynamics of the universe, scalar field whose potential is subject to some amount of fine-tuning (arguably at the $10^{-12}$ level),  it usually implies the existence of a singularity in the far past, may face a trans-Planckian problem, and demands that quasiclassical General Relativity (GR) be valid up to energies of the order $10^{-3}\mP$, with $\mP^{-2}\equiv \GN$ the Planck mass ($\GN$ being Newton constant).  Of these problems, none is by itself sufficient to reject the paradigm, but the combination might provide a good motivation either to use a different approach or to complement inflation somehow.

The pre Big-Bang scenario (see the review~\cite{GV03}) was the first to set the issue along these lines, followed, after quite some time, by an application of the brane idea in \eg Ref.~\cite{Shtanov:2002mb} and in the so-called ``ekpyrotic'' model (see for instance~\cite{KOST01,KKL01}).  All these models\footnote{We do not here consider the other category of models based on string gas cosmology~\cite{BFK071, BFK072,NBV06, BNPV071}, as they do not contain a bounce phase.}, written in the Einstein frame, contain a contracting phase preceding the ongoing expanding one.  The reversal between these phases yields a bounce in the scale factor~\cite{Tolman31,Starobinsky78,DL96,Kanekar:2001qd}, with an impact on cosmological observations for which general properties have not yet been obtained. Note that roughly at the same time as these models were developed, discussions based purely on GR had concluded that no bounce could be achieved with a perfect fluid source~\cite{PP02}.

Among previous works on bouncing cosmologies, those that stay within the framework of 4-dimensional GR~\cite{MP03,GGV04} have focused solely on the bounce-producing mechanism. With a single dynamical degree of freedom, a scalar field say, this requires closed spatial sections in order to satisfy the null energy condition all along.   Other choices are possibles.   Achieving a bounce in 4 dimensions with flat or even open spatial sections requires one to violate the null energy condition and the presence of at least two kinds of fluids, one of which must have negative energy density~\cite{PP022,GGV03,BV05,AW04,Cai+07}.  As another alternative, a scalar field with a nonconventional kinetic term can be used, the so-called  ``K$-$bounce''~\cite{AP07}.  There also exists phenomenological descriptions which aim to classify the possible perturbation evolution through the bounce~\cite{ABB07,BM06}.

Having managed a background bounce, one needs to describe the evolution of scalar and tensor perturbations in this background for comparison with observations (note also the existence of other possible tests of bouncing scenarios, \eg Ref.~\cite{SlGlKG05}). Propagating perturbations through the bounce itself is very intricate~\cite{MP041, MP042} as one wants perturbations to remain small at all times. Unfortunately, Einstein gravity is often spoiled by the necessarily weird conditions to be imposed in order for a bounce to actually take place. In particular, having a negative energy fluid clearly leads to instabilities in the long run; this is especially true for models having an infinitely long contracting phase. This can however be handled by assuming the bounce-making negative energy component to be merely phenomenological, acting only for a short time, the bounce duration, in general assumed comparable with the Planck scale. In any case, there are no generic properties that can be derived in a model-independent way as in slow-roll inflation~\cite{MR07}. Very often, though, it is found that a reasonable model connects two almost de Sitter phases.

The presence of an expanding de Sitter phase after the bounce naturally leads to the question of whether one can implement an inflationary epoch connected to a bounce phase, in a mixed situation somehow reminiscent of the topological defects~\cite{VS00} versus inflation challenge: as topological defects could not be identified as the sole source of primordial perturbations, in particular because of their incoherent spectrum, it was suggested to switch the question from ``inflation or defects'' to ``inflation and defects or inflation only''~\cite{RS05}, especially in view of the fact that most reasonable grand unified theories~\cite{Langacker81} are expected to produce such defects at the end of inflation~\cite{JRS03}. The idea of the present work relies on the same point of view: if a purely bouncing alternative is found to be unrealistic, why not consider a situation in which inflation follows a bounce?

The purpose of this work is thus to study a class of models implementing this idea. We assume a bouncing phase, obtained by a simple scalar field evolving in a universe with closed spatial section, followed by an inflationary epoch. This second era then provides the standard solution for some usual cosmological issues (flatness and homogeneity as well as reheating), while the bounce permits the model to avoid a primordial singularity and provides an infinite horizon. The hope is then to find out whether any effect could somehow be present in the primordial spectrum, leaving some imprint to be tested against observations.

In the following section, \S~\ref{sec:background}, we discuss the background model and recall the necessary conditions for the occurence of a bounce; this section is heavily based on Ref.~\cite{MP03}. We also implement constraints that were not considered in this reference and characterize the potential through which perturbations propagate (this is detailed in \S~\ref{sec:perturbations}). \S~\ref{sec:phase} is then devoted to describing how a bounce followed by inflation can take place in our framework by means of a phase space analysis.  This analysis indicates that, as expected, some amount of fine-tuning is demanded.  In \S~\ref{sec:perturbations} we obtain the potential through which perturbations propagate and describe the propagation of scalar perturbations in this background. We derive the resulting spectrum, which turns out to be the product of an oscillatory component and an almost flat power-law component at leading order, and find that the curvature perturbation $\zeta_\mathrm{BST}$ (as first defined by Bardeen, Steinhardt and Turner~\cite{BST83}) is not conserved through the bounce.  On the one hand, we can conclude that, based on the current belief that the spectrum of primordial perturbations at horizon exit should be scale-invariant and almost flat with possibly superimposed features at higher order, this class of models can most likely be generically ruled out unless a justification is found under which the wavelength of the oscillations is sufficiently long so as to make them unnoticeable.  On the other hand, because the cosmic microwave background (CMB) multipoles $C_{\ell}$ are roughly given by the convolution of a spherical Bessel function with the initial power spectrum we find that evolving this spectrum through the radiation and matter dominated eras damps the oscillations and results in a spectrum of CMB multipoles not incompatible with the Wilkinson Microwave Anisotropy Probe (WMAP) data~\cite{Dunkley+08,Komatsu+08}.
\section{Background bouncing model} \label{sec:background}

Our starting point consists of GR with a scalar field, \ie we assume the dynamics to derive from the action
\be
\mathcal{S} = \int\dd^4x\sqrt{-g}\left[ \frac{R}{6\lP^2} -\frac{1}{2}\partial_\mu\varphi
\partial^\mu\varphi -V\left(\varphi\right)\right],
\label{eq:action}
\ee
with $\lP^2=\frac{8}{3}\pi\GN$ the Planck length, ($\GN$ being Newton constant) and $V\left(\varphi\right)$ the self-interaction potential for the scalar field $\varphi$, which is left unspecified for the time being. In what follows, for notational convenience, we shall assume the natural system of units in which $8\pi\GN=1$.  Assuming homogeneity and isotropy, the background which solves the equations of motion derived from Eq. (\ref{eq:action}) takes the Friedmann-Lema\^{\i}tre-Robertson-Walker form and reads
\be
\mathrm{d}s^2=\mathrm{d}t^2-a(t)^2\left(\frac{\mathrm{d}r^2}{1-\Ka r^2}+
r^2\mathrm{d}\theta^2+r^2\mathrm{sin}^2\theta\mathrm{d}\phi^2\right),
\label{metric}
\ee
where $a(t)$ is the scale factor, and where the spatial curvature $\Ka>0$ can be normalized to unity. The scalar field $\varphi$ can be understood as either fundamental or phenomenological, but in both cases, its energy density $\rho_{\varphi}$ and pressure $p_{\varphi}$ are given by
\ba
\rho_{\varphi}=\frac{\dot {\varphi}^{2}}{2}+V(\varphi),\quad p_{\varphi}=\frac{\dot{\varphi}^{2}}{2}-V(\varphi).
\label{DensityPressure}
\ea
In Eq.~(\ref{DensityPressure}) and in the following, a dot denotes a derivative with respect to cosmic time $t$. In the forthcoming calculations, we also use the conformal time $\eta$ defined through
\be
\mathrm{d}t = a(\eta) \dd\eta,
\label{eq:eta}
\ee
and unless specified otherwise, derivatives w.r.t. $\eta$ will be denoted by a prime, so that for an arbitrary function of time $f$,
one has $f' = a \dot f$.

Einstein's field equations relate the time evolution of the scale factor $a(t)$ to the stress tensor for $\varphi$ (\ie to the pressure $p_{\varphi}$ and the energy density $\rho_{\varphi}$) through
\ba
&\displaystyle H^2 = \displaystyle \frac{1}{3}\left(\frac{1}{2}\dot{\varphi}^2+V\right)-\frac{\Ka}{a^2}
\label{Friedmann1},\\
&\displaystyle \dot{H}  =\displaystyle  \frac{1}{3}\left(V-\dot{\varphi}^2\right)-H^2
\label{Friedmann2},
\ea
where $H\equiv\dot{a}/a$ is the Hubble expansion rate, while the (redundant) dynamical Klein-Gordon equation for $\varphi$ is given by
\be
\displaystyle \ddot{\varphi}+3H\dot{\varphi}+V_{,\varphi}=0.
\label{KleinGordon}
\ee
Combining Eqs.~(\ref{Friedmann1}) and (\ref{Friedmann2}) yields the following relations
\ba
\displaystyle \rho_{\varphi}+p_{\varphi} & = & 2\left(\frac{\Ka}{a^{2}}-\dot{H}\right),\\
\displaystyle \rho_{\varphi}+3p_{\varphi} & = & -6\left(\dot{H}+H^2\right),
\ea
with which the energy conditions can be rephrased in terms of the time behaviour of the Hubble rate $H$ and scale factor $a$. Since, by definition, one has $H=0$ and $\dot{H}=\ddot{a}/a_0>0$ at the bounce, the null energy condition (\ie, $\rho_{\varphi}+p_{\varphi}\geq 0$) is only preserved provided $\Ka>0$, hence the choice of positively curved spatial sections. On the other hand, as in inflationary scenarios, the strong energy condition (namely $\rho_{\varphi}+3p_{\varphi}\leq0$) is necessarily violated.  Note that this violation of the strong energy condition is nothing but the usual requirement that $\ddot{a}>0$ in an inflationary stage.

Following Ref.~\cite{MP03}, we now switch to a description in terms of conformal time as defined above and expand the scale factor $a$, the scalar field $\varphi$ and its potential $V(\varphi)$ around the bounce, set, for definiteness, to take place at $\eta=0$. To fourth order in $\eta$, we have
\begin{widetext}
\ba
a(\eta) & = & a_{0}\left[1+\frac{1}{2}\left(\frac{\eta}{\eta_{0}}\right)^{2} +\frac{\delta}{3!}\left(\frac{\eta}{\eta_{0}}\right)^{3}+\frac{5}{4!}\left(1+\xi \right)\left(\frac{\eta}{\eta_{0}}\right)^{4}\right]+\mathcal{O}\left(\eta^{5}\right),\label{a1}\\
\varphi(\eta) & = & \varphi_{0}+\varphi'_{0} \eta+\frac{1}{2}\varphi''_{0} \eta^{2}+\frac{1}{3!}\varphi'''_{0} \eta^{3}+\frac{1}{4!}\varphi^{IV}_{0} \eta^{4}+\mathcal{O}\left(\eta^{5}\right)\label{phi}\\
V(\varphi) & = & V(\varphi_{0})+V_{,\varphi} \varphi'_{0}\eta+\frac{1}{2}\left(\frac{\dd^{2}V}{\dd\varphi^{2}} \varphi'^{2}_{0}+\frac{\dd V}{\dd\varphi} \varphi''_{0}\right)\eta^{2}+\mathcal{O}(\eta^{3})\label{V},
\ea
\end{widetext}
where $\eta_0$ ($a_0 \eta_0$ in physical units) defines the duration of the bounce and provides a natural time scale. In Eq.~(\ref{a1}), the fourth order term with $\xi \ne 0$, is written in this way in order to emphasize the deviation from the quasi-de Sitter solution, namely $a(\eta)=a_0\sqrt{1+\mathrm{tan}^2\left(\eta/\eta_0\right)}$; this is discussed in Ref.~\cite{MP03}. The asymmetry parameter in the $\mathcal{O}\left(\eta^3\right)$ term with $\delta \ne 0$ thus corresponds to a deviation from a purely symmetric bounce.

Inserting Eqs.~(\ref{a1}), (\ref{phi}) and (\ref{V}) into Eqs.~(\ref{Friedmann1}), (\ref{Friedmann2}) and (\ref{KleinGordon}), and assuming\footnote{In the case of a symmetric bounce, $\varphi'_\star=0$ corresponds to a de Sitter bounce since \begin{displaymath} \lim_{\varphi'_0\to 0}\left(\rho_{\varphi}+p_{\varphi}\right) =0 \end{displaymath} and in the case of an asymmetric bounce, there is singular behaviour in the evolution of the perturbations.} $\varphi'_0 \ne 0$, one can relate the scale factor parameters to the field values at the bounce as follows,
\be
a_{0}^{2} = \frac{3-\Upsilon}{ V_0},
\label{a2}\ee
which gives the value of the scale factor at the bounce, and where $V_0 \equiv V(\varphi_0)$ and $\Upsilon\equiv \frac12 \varphi_0'^2$, as in Ref.~\cite{MP03}, 
\be \eta_0^2  = \frac{1}{1-\Upsilon},
\label{eta0}\ee
which gives the characteristic duration of the bounce, 
\be
\delta  =  \frac{\left(3-\Upsilon\right)\sqrt{\Upsilon}}{3\sqrt{2}
\left(1-\Upsilon\right)^{\frac{3}{2}}}\frac{V'_0}{V_0},
\label{delta}
\ee
and 
\ba
\xi & = & \frac{4 \Upsilon (\Upsilon-1)}{5 (\Upsilon-1)^2}-\left(\frac{V_0'}{V_0}\right)^2 \frac{(\Upsilon-3)^2}{5   (\Upsilon-1)^2}+\label{xi} \\ \nonumber
& &\frac{V_0''}{V_0}\frac{2 \Upsilon (3-\Upsilon)}{5   (\Upsilon-1)^2},
\ea
the two parameters describing the deviation from a purely de Sitter bounce.  
Note that in Eqs.~(\ref{delta}) and (\ref{xi}), we have defined a prime, when applied to the potential $V$, to denote a derivative with respect to $\varphi$, the function being evaluated at the bounce, {\ie $V'_0\equiv V_{,\varphi}(\varphi_0)$}. Note that since $\eta_{0}^{2} \geq 1$, we also  have $0 \leq \Upsilon \leq 1$.

As made clear in~\cite{MP03}, if one restricts the analysis to a symmetric bounce, one has $\delta=0$ such that either $\varphi'_0=0$, in which case $\Upsilon=0$, or $\varphi_0''=V_{,\varphi}=0$. The first case, $\varphi'_0=0$, corresponds to the exact de Sitter case ($\eta_0=1$): the null energy condition is then only marginally violated and only the gauge modes of scalar perturbations interact with the potential term (see Ref.~\cite{GT03} for a detailed examination). The second situation, having $\Upsilon\not=0$,  demands that both $\varphi_0''$ and $V_{,\varphi}$ vanish at the bounce.

In both cases, one has $\displaystyle \varphi''=\varphi'' \left[V\left(\varphi\right),\varphi',\mathcal{H}\right]$ so that requiring a symmetric bounce shrinks the continuous set of all possible trajectories to a discrete set, \ie in a symmetric bounce, the solutions of Friedmann's equations at the bounce are a denumerable set.

Another interesting qualitative result can be obtained by expanding the kinetic term $\frac12\dot{\varphi}^2$ around the bounce.  One gets
\ba
\displaystyle \frac{\varphi'^2}{2} & = & \Upsilon+\frac{V'_0}{V_0}\sqrt{2}\sqrt{\Upsilon \left(\Upsilon-3\right)}\eta+\nonumber\\
& & \displaystyle \left[2 \Upsilon\left(1-\Upsilon\right)+2 \left(\frac{V'_0}{V_0}\right)^2 
\left(\frac{\Upsilon}{2}-3\right)^2+\right.\nonumber\\
& &
\displaystyle \left.
\frac{V''_0}{V_0} \Upsilon \left(\Upsilon-3\right)\right]\eta^2,
\ea
the right-hand side of which must evidently remain positive for $\varphi$ real.  In the (quasi)symmetric case for which one can ignore all $V'_0$ terms, this simplifies to
\be
\frac{1}{2}\frac{V''_0}{V_0} \frac{\Upsilon-1}{\Upsilon-3} \leq 0.
\ee
Given the bounds on $\Upsilon$, this means that $V''_0\leq 0$: the potential must have a convex part, which the field explores right at the very moment of the bounce. This restricts the possible shapes for the self-interacting potential $V\left(\varphi\right)$; our specific choice of Eq.~(\ref{V2}) in \S \ref{sec:phase} does indeed fulfill such a requirement. Note that in the more generic asymmetric case, although we believe that in small deviations from the symmetric case, the convex shape of $V\left(\varphi\right)$ at the time of the bounce remains necessary, no such firm restriction can in fact be obtained, and more freedom in the choice of $V\left(\varphi\right)$ is expected to be allowed.

In the case of a general asymmetric bounce, it is worth noting that there are two very distinct ways in which $\varphi$ may evolve if the condition $\varphi' \ne 0$ imposed in Eqs.~(\ref{a2}) to (\ref{xi}) is relaxed. One may obtain such a bounce if $\varphi$ either moves up the potential until $\varphi'=0$, at which point it returns towards the value it started from or if it evolves from one minimum of the potential to another one.  As we shall see, the former implies singular behaviour of the perturbation equations at the bounce, and so is not considered any further in the subsequent analysis while in the latter case, one necessarily has $\varphi_0' \ne 0$, $\varphi_0''\ne 0$ and $V_{,\varphi} \ne 0$, and the set of solutions is a finite volume of phase space dependent on the form of $V\left(\varphi\right)$ only. We discuss these points further in the following section.

\section{Phase Space Analysis}
\label{sec:phase}

In the previous section, we were able to determine some of the properties $V\left(\varphi\right)$ should satisfy for a bounce to occur.  In this section, further insight into the full dynamics is gained by means of a phase space analysis of the evolution of both the Hubble parameter and the scalar field in the $\left( \varphi,\dot{\varphi},H \right)$ phase volume.  Such an analysis is possible only once a specific form of the potential $V(\varphi)$ is given.  In the remainder of the paper, we consider a spontaneously broken symmetry (Mexican-hat) potential of the form
\be
V(\varphi)=V_0-\frac{\mu^2}{2}\varphi^2+\frac{\lambda}{4!}\varphi^4.
\label{V2}
\ee
where $\mu$ and $\lambda$ are the mass and self-interaction parameters respectively, and $V_0$ is the height of $V(\varphi)$ for $\varphi=0$. Such a potential, quite apart from being often used for different inflationary models, is theoretically well motivated for instance as stemming from a grand unified theories framework, in which case $\varphi$ is identified with a component of a larger scalar multiplet, the other degrees of freedom of which are somehow frozen at the energy scales under consideration.

The potential (\ref{V2}) might lead to the formation of domain walls~\cite{Kibble76,Vilenkin85}, which, forming in the very early universe, would spoil its overall evolution and result in a singular crunch~\cite{Everett74}. However, if $\varphi$ belonged to a larger multiplet, the symmetry-breaking scheme would hopefully be sufficiently different so as not to produce such defects, although even if the decoupling of the various degrees of freedom is effective enough that domain walls should have formed, it is not absolutely clear whether it would have such a dramatic impact. Consider first the inflationary case. The maximum temperature, once thermalization is taken into account, is found to be quite below the Hagedorn temperature, so that the symmetry is, in fact, not restored and no defects can be formed at all as one proceeds backwards in time~\cite{EE90,ES93}. In the bouncing case for which one starts out with a large and cold universe, the argument does not stand anymore, as the field must be assigned arbitrary values in regions of spacetime separated by a distance larger than its correlation length, in practice its Compton wavelength (note that the ``causality'' argument~\cite{Kibble76} cannot apply in this case, as the horizon can be assumed infinite in the past infinity at which one sets the initial conditions). Taking into account the effects caused by the evolution of the background, it seems probable, though by no means guaranteed, that the subsequent contraction would result in high decay rates of wall and antiwall configurations, possibly leading to a simple and automatic solution of the wall puzzle (with tremendous production of particles).  The remaining wall distribution can, once transfered into the expanding epoch, be made consistent with observations, \eg the CMB~\cite{Dunkley+08,Komatsu+08}, provided the network is strongly frustrated~\cite{FMP03}, or if an initial bias, however tiny~\cite{LSW97,CS02,Hindmarsh96}, is introduced in the potential.
\begin{figure*}
\begin{center}
\includegraphics[width=5.75cm]{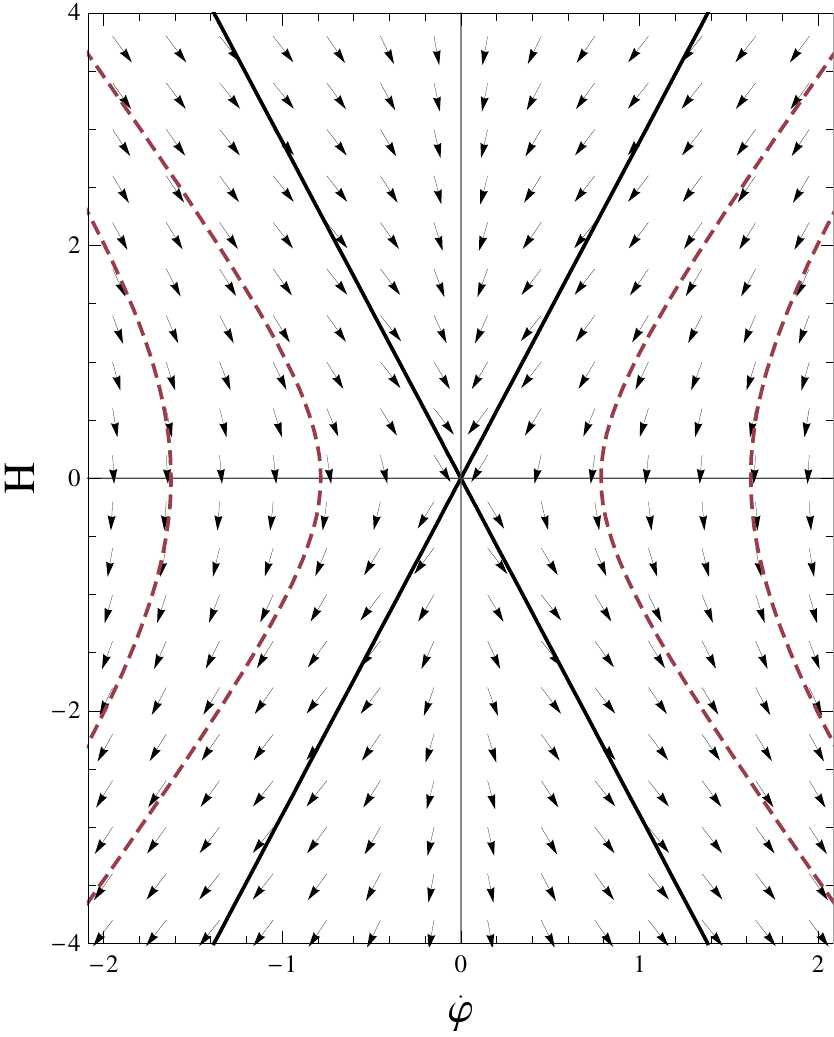}
\hspace{0.1cm}
\includegraphics[width=5.75cm]{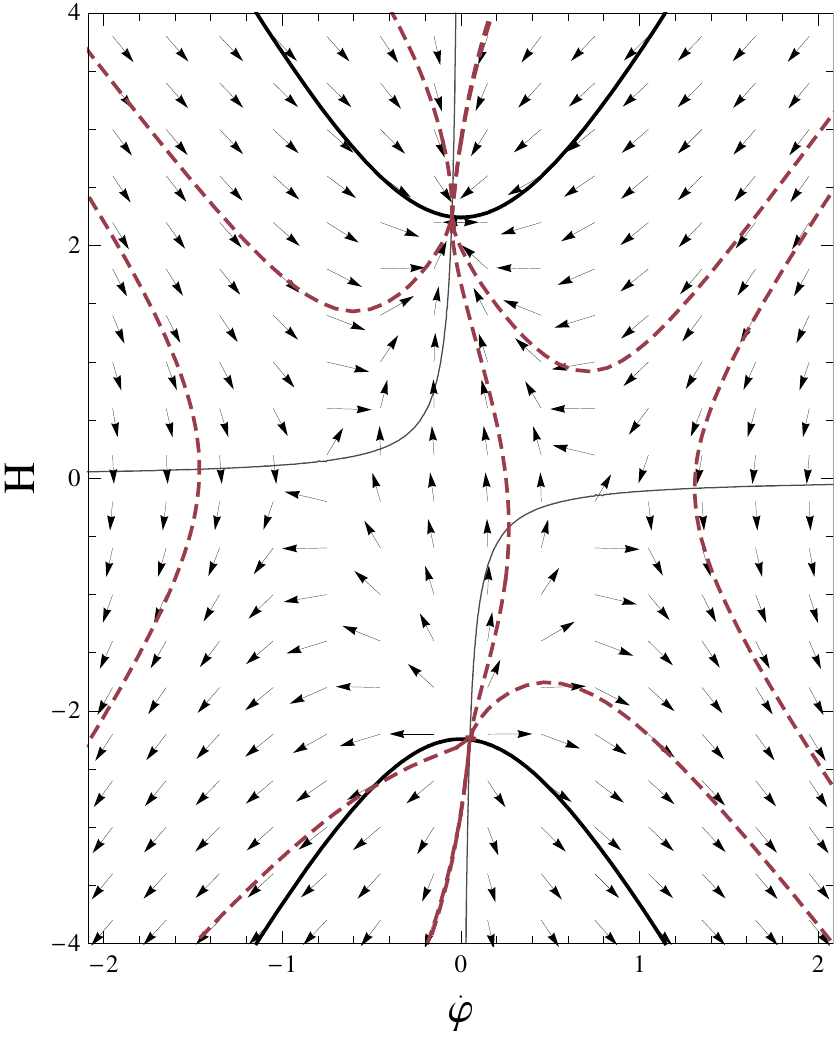}
\hspace{0.1cm}
\includegraphics[width=5.75cm]{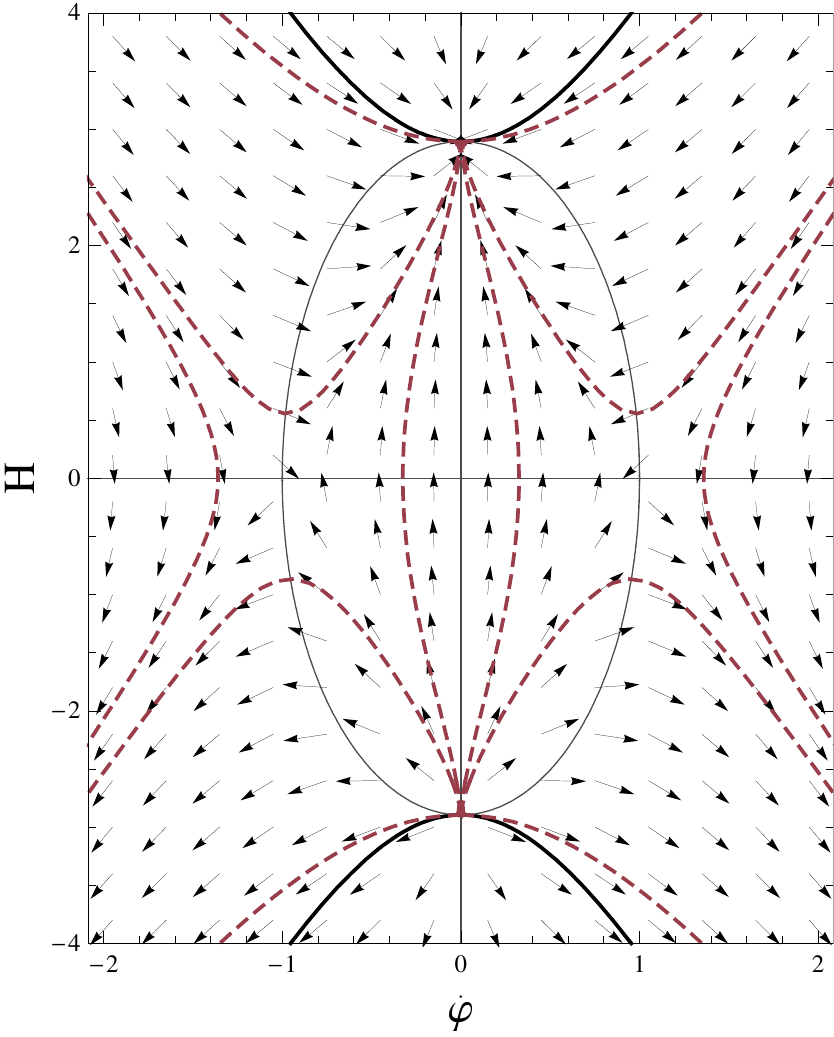}
\end{center}
\caption{Phase portrait in the $\left( \dot{\varphi},H \right)$ plane for $\mu=3/2$, $\lambda=3\mu^4/2 V_0$ and fixed values of $\varphi$, equal to $-2$, $-1.25$, $0$ in the left, center and right panels respectively.  The flow vectors are located at equally spaced values of $\dot{\varphi}$ and $H$, are normalized to magnitude unity and their direction is determined using (\ref{Friedmann2}) and (\ref{KleinGordon}). In each panel the isoclines are indicated by thin lines and the thick parabolic curves are the $\mathcal{K}=0$ solutions, \ie $H=\pm \sqrt{\frac{1}{3}\left(\dot{\varphi}^2+V\right)}$.  The region in between the curves corresponds to $\mathcal{K}=1$, while the two remaining regions, one of which is above the top thick curve and the other below the bottom thick curve, are $\mathcal{K}=-1$ regions.  The trajectories displayed as thick dashed lines correspond to solutions of the coupled system (\ref{Friedmann2}) and (\ref{KleinGordon}), some of which connect a slow-roll ($\dot{\varphi}\ll 1$), $H<0$ region across $H=0$ onto a slow-roll ($\dot{\varphi}\ll 1$), $H>0$ region, while others are either singular or purely inflationary solutions in the region $\mathcal{K}=-1$.  In the center panel increasingly more trajectories have singular behavior as compared to the panel on the right because the region containing nonsingular bouncing solutions shrinks as one moves away from $\varphi=0$, \ie away from the top of the potential $V \left( \varphi \right)$.  In the left panel, which corresponds to the bottom of the potential $V \left( \varphi \right)$, this region has shrunk down to a single point.}
\label{PhaseSpace1}
\end{figure*}

As it turns out, the phase space analysis is most easily performed using the cosmic time variable $t$. It provides some valuable insight, particularly on the space of initial conditions that ensure the stability of the background spacetime and the occurence of a bounce somewhere in its history.  As we shall see, a potential of the form (\ref{V2}) leads naturally from a contracting phase to a bounce and on to a slow-roll type inflationary phase.  In what follows, we first discuss the set of possible trajectories restricting ourselves to the planes $\left(H,\dot{\varphi}\right)$ and $\left(\varphi,\dot{\varphi}\right)$ and then go on to describe trajectories in the full volume $\left(H,\varphi,\dot{\varphi} \right)$.
\begin{figure}[h]
\includegraphics[width=7.5cm]{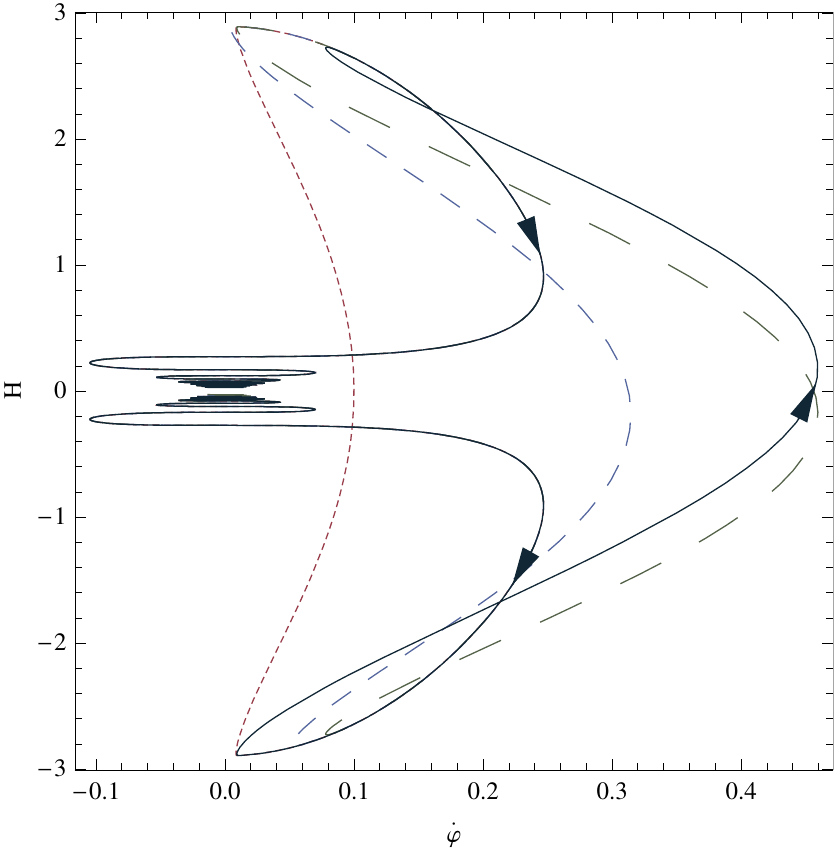}
\caption{Trajectories projected in the $\left( \dot \varphi,H \right)$ plane for $\mu=3/2$, $\lambda=3 \mu^4/2V_0$ and for four different sets of initial conditions on $\left( H,\varphi,\dot{\varphi} \right)$, namely  $\left( 0, 0, 1/10 \right)$ [dotted],  $\left( 7/10, 0, 1/10 \right)$ [short-dashed], $\left( 3/5 \sqrt{V_0/3},0,1/5 \right)$ [long-dashed] and $\left( -3/5 \sqrt{V_0/3},0,1/5 \right)$ [full]. Note that trajectories are practically undistinguishable on either side of the short bouncing phase during which they cross the $H=0$ axis.}
\label{traj1}
\end{figure}
To build the phase portrait in the ($\dot{\varphi}$,$H$) plane, we start with a set of evenly distributed initial values for $\dot{\varphi}$ and $H$, for a fixed value of $\varphi$, and construct flow vectors with components $\dot{H}$ and $\ddot{\varphi}$ defined using (\ref{Friedmann2}) and (\ref{KleinGordon}) as illustrated on Fig.~\ref{PhaseSpace1}.  The $\Ka=+1$ region is bounded above and below by the $\Ka=0$ solution $H=\pm \sqrt{\frac{1}{3}\left(\dot{\varphi}^2+V\right)}$, denoted by the two parabolic thick curves in Fig.~\ref{PhaseSpace1}, beyond which the trajectories describe the $\Ka=-1$ case. A simplified view of the full dynamics, as provided by the three slices of the figure, readily indicates that a closed universe undergoing a period of slow-roll type inflation has a nonzero probability of having crossed the $H=0$ plane in the past and therefore undergone a period of slow-roll type contraction prior to the bounce, as indicated by the two central trajectories (denoted by the dashed curves) in the right panel.  Projections of typical bouncing trajectories in the $\left(H,\dot{\varphi}\right)$ plane are shown in Fig.~\ref{traj1} for various initial conditions.  As we shall see in the subsequent analysis, there exists, in the full 3D phase space, two saddle points located at $H \simeq \pm 3$ and $\varphi=\dot{\varphi}\simeq 0$. In the 2D slices of Fig.~\ref{PhaseSpace1}, these saddle points reduce to a repulsor and an attractor respectively, not only in the $\varphi=0$ slice but also in the two other ones, for which the locations of the critical points of course shift away from $H \simeq \pm 3$ to $\simeq \pm 1.75$ (center panel) and then collapse into a single point (left panel).  Given the existence of these two saddle points, some trajectories describe a situation in which a universe undergoing slow-roll contraction naturally bounces and then experiences slow-roll inflation in the usual way.  Going from right to left in the figure, in the direction of increasing $\varphi$, one finds that the stable region in which a bounce occurs shrinks and the left-right symmetry of the right panel is lost.  When the field reaches the bottom of $V \left( \varphi \right)$, $V_{\mathrm{min}}$, the region has shrunk to a point.  Some other trajectories, on the other hand illustrate a singular evolution for which $H$ grows increasingly negative and $\varphi$ is driven to $\infty$.

\begin{figure*}
\begin{center}
\includegraphics[width=7cm]{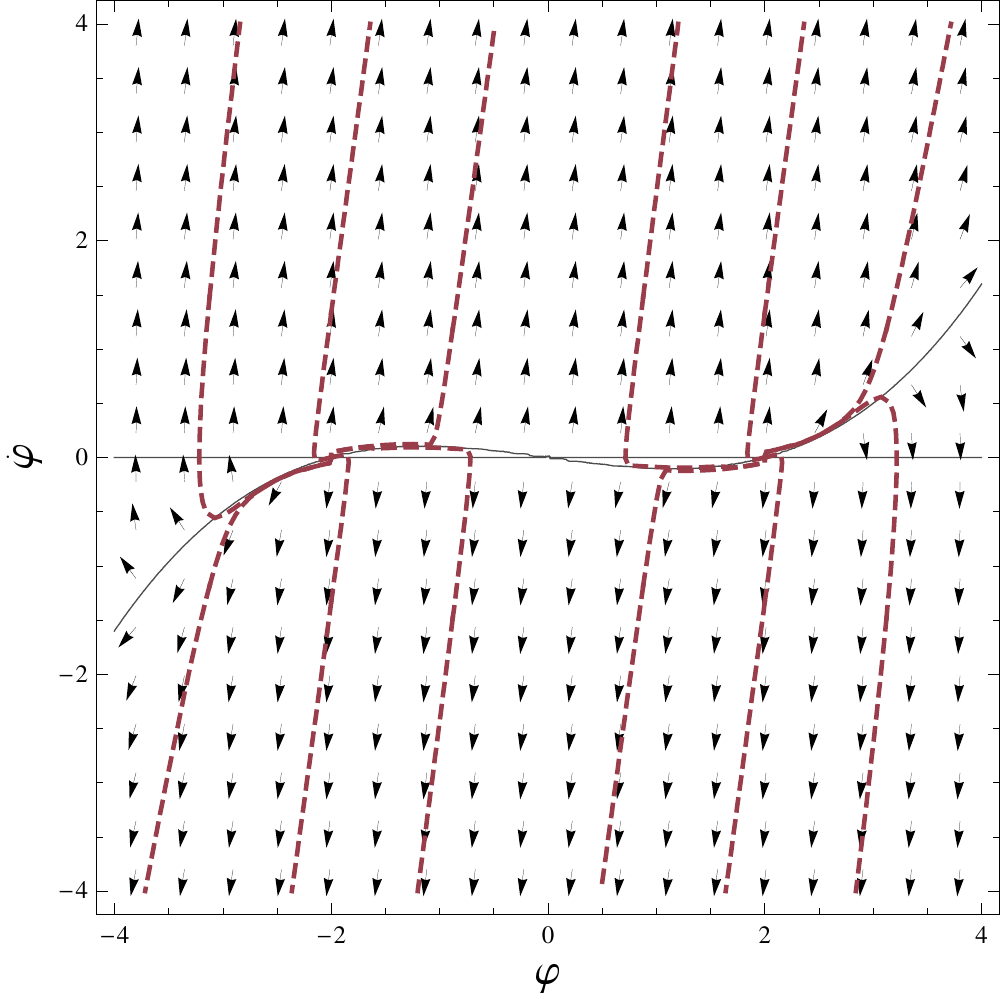}
\hspace{0.3cm}
\includegraphics[width=7cm]{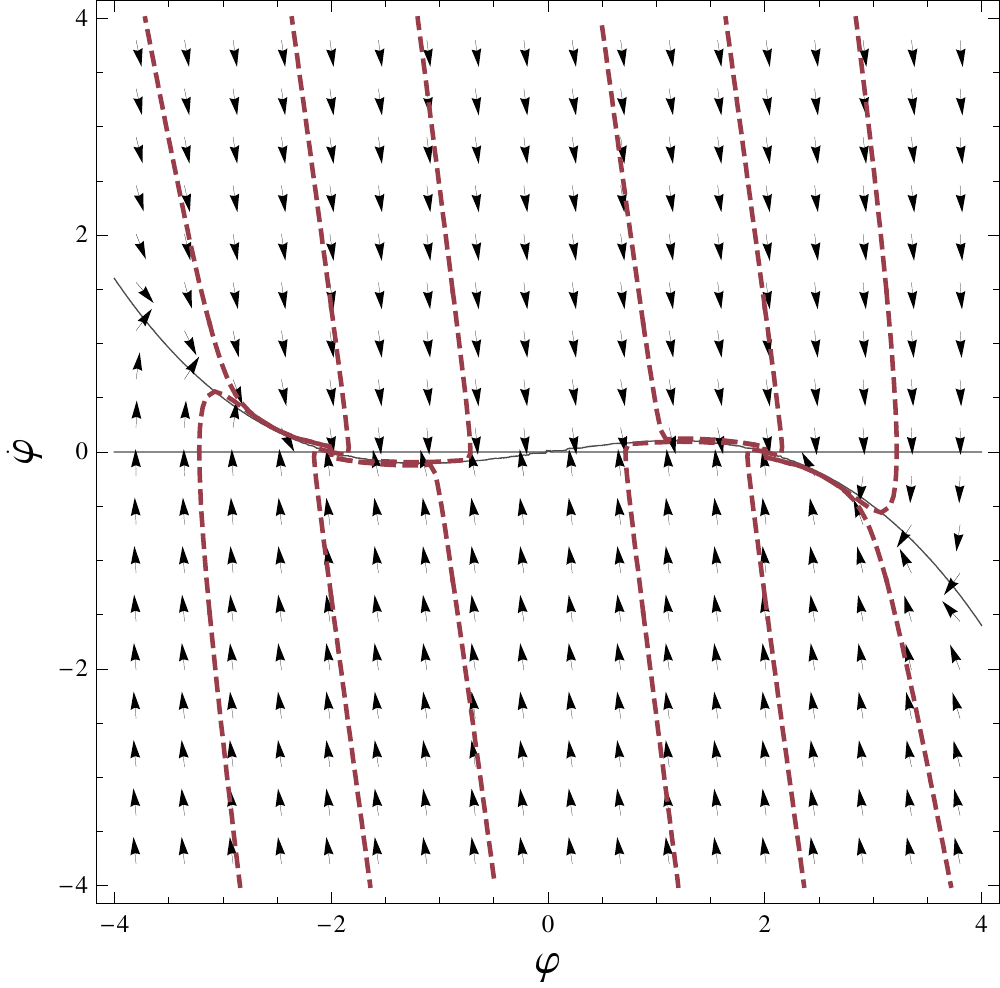}
\includegraphics[width=7cm]{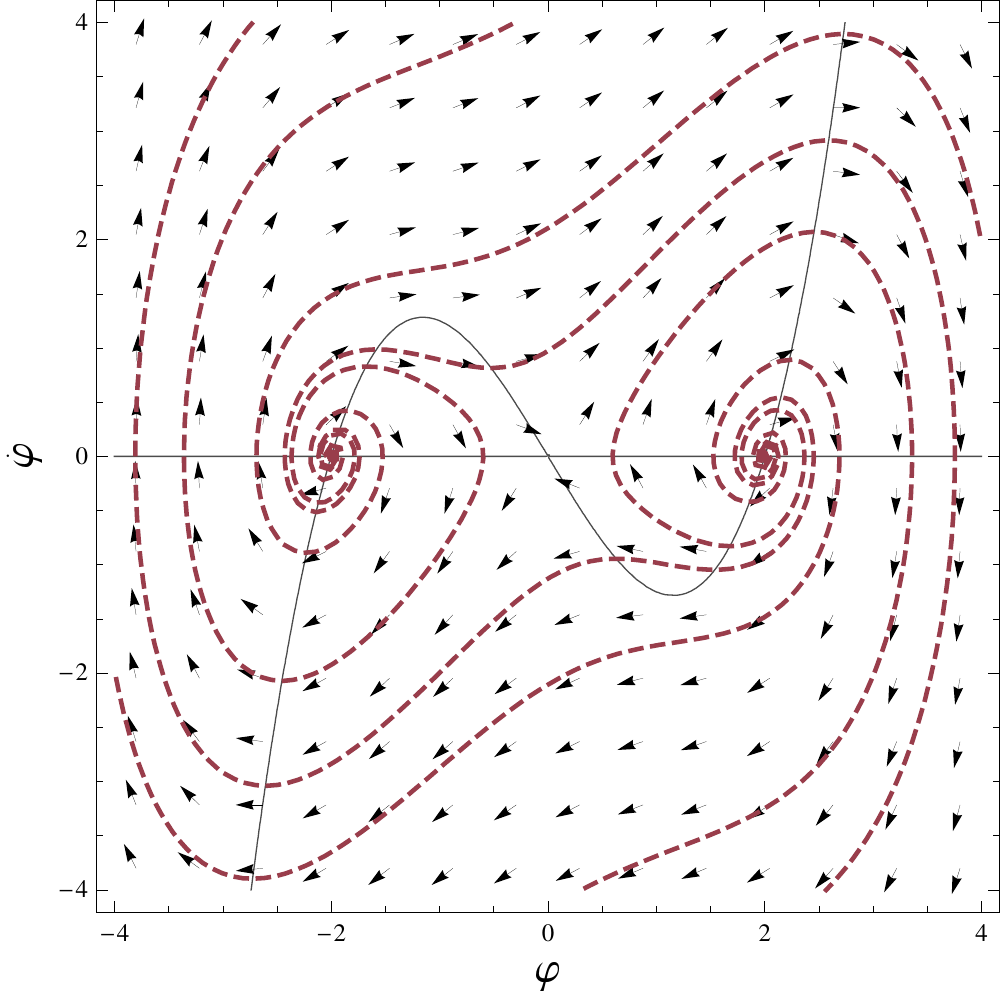}
\hspace{0.3cm}
\includegraphics[width=7cm]{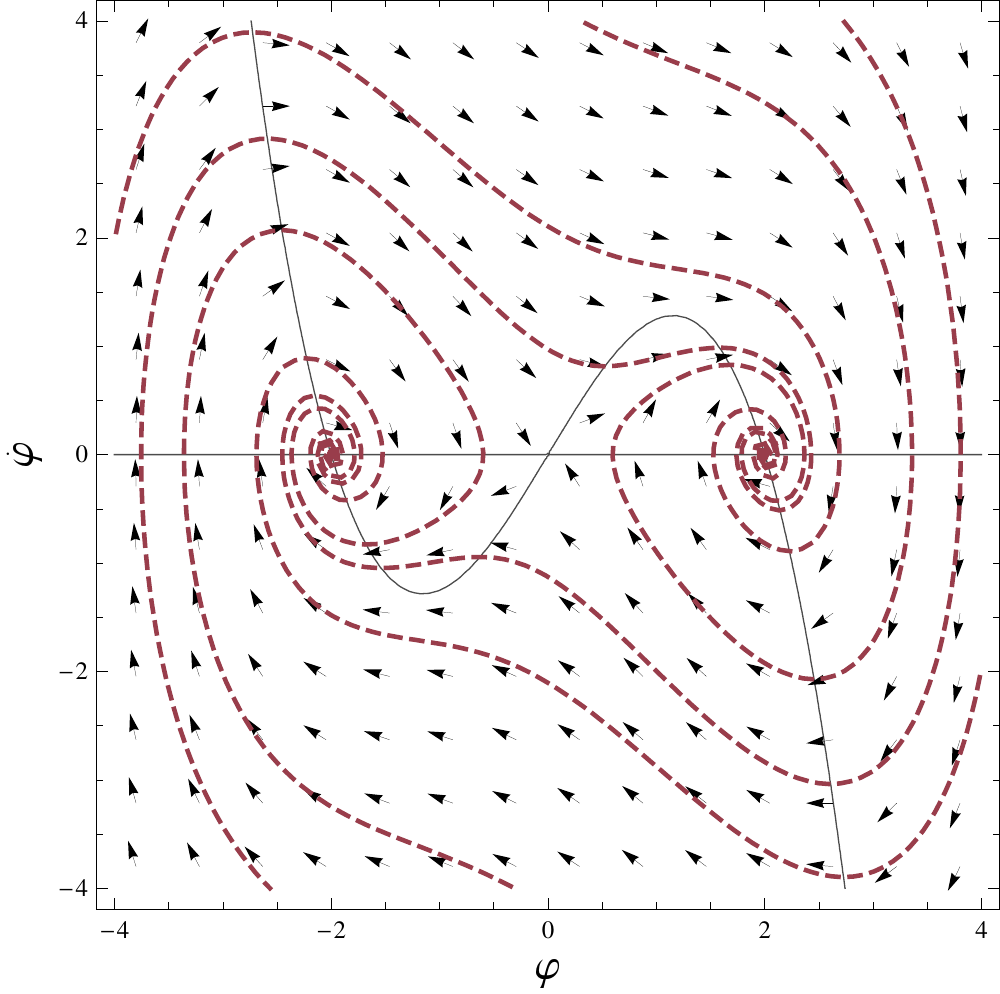}
\end{center}
\caption{Phase portrait in the ($\varphi$,$\dot{\varphi}$) plane for $\mu=3/2$, $\lambda=3\mu^4/2 V_0$ and for fixed values of the Hubble parameter $H$, equal to $-\sqrt{V_0/3}\simeq -3$, $\sqrt{V_0/3}\simeq 3$ in the top left and right panels respectively and $-0.1$, $0.1$ in the bottom left and right panels respectively.  When $H<0$ ( left-hand side), the system is naturally pushed away from $\varphi=-2$ towards large values of both $\varphi$ and $\dot{\varphi}$.  The opposite is true when $H>0$ ( right-hand side).  Combined with Fig.~\ref{PhaseSpace1}, it is clear that initial conditions taken at the bottom of the potential, \ie $\varphi \simeq -2$, $\dot{\varphi} \ne 0$ and $H \le 0$, have to be appropriately (fine) tuned in order for the system to stay nonsingular (\ie $\varphi$ and $\dot{\varphi}$ remaining small) and to evolve towards the region in which a bounce ensues.  When $H>0$ (top and bottom right) the system remains stable and naturally tends towards the bottom of the potential where $\varphi=2$.}
\label{PhaseSpace2}
\end{figure*}
\begin{figure*}
\begin{center}
\includegraphics[width=11.5cm]{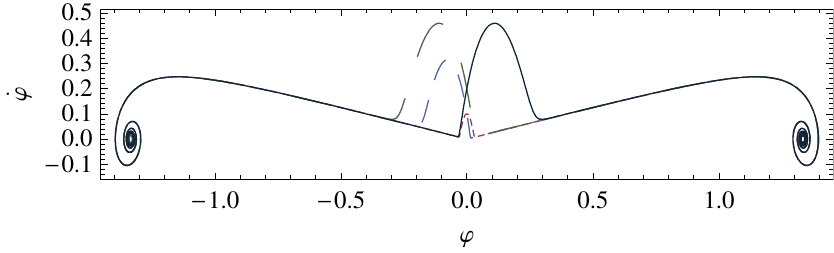}
\end{center}
\caption{Trajectories projected in the $\left( \varphi,\dot{\varphi} \right)$ plane with the same sets of parameters as in FIG.~\ref{traj1}.}
\label{traj2}
\end{figure*}

The phase portrait in the ($\varphi$,$\dot{\varphi}$) plane can be constructed in a similar way.  To do so, we fix $H$ and set up $\varphi$ and $\dot{\varphi}$ initial values with a flow given by the $\dot{\varphi}$ values as the vector component in the $\varphi$ direction, and use Eq.~(\ref{KleinGordon}) to define components in the $\dot{\varphi}$ direction (see Fig.~\ref{PhaseSpace2} and Fig.~\ref{traj2}).  On the top and bottom left panels of Fig.~\ref{PhaseSpace2}, $H<0$, and if $\varphi$ is perturbed away from $V_{\mathrm{min}}$, both $\varphi$ and $\dot{\varphi}$ will be driven to large values.  The top and bottom right panels show that for $H>0$, $\varphi \rightarrow \varphi_+\simeq2$ ($\varphi_\mp$ are the values of $\varphi$ for which $V\left( \varphi \right)$ is at its minimum $V_{min}$) and $\dot{\varphi} \rightarrow 0$, as expected.  Note also that as $H \rightarrow 0^{\pm}$, trajectories are increasingly circular and that in each 2D slice, there exists spiral points at $\varphi=\pm2$ and $\dot{\varphi}=0$.  Anticipating on the forthcoming results of the 3D phase space analysis, one notes that by arranging the initial conditions for $H$ and $\dot{\varphi}$ appropriately near $V_{\mathrm{min}}$, $\varphi$ can be pushed towards zero, the region in which trajectories are nonsingular and undergo a bounce will be approached.

\begin{figure}
\includegraphics[width=8cm]{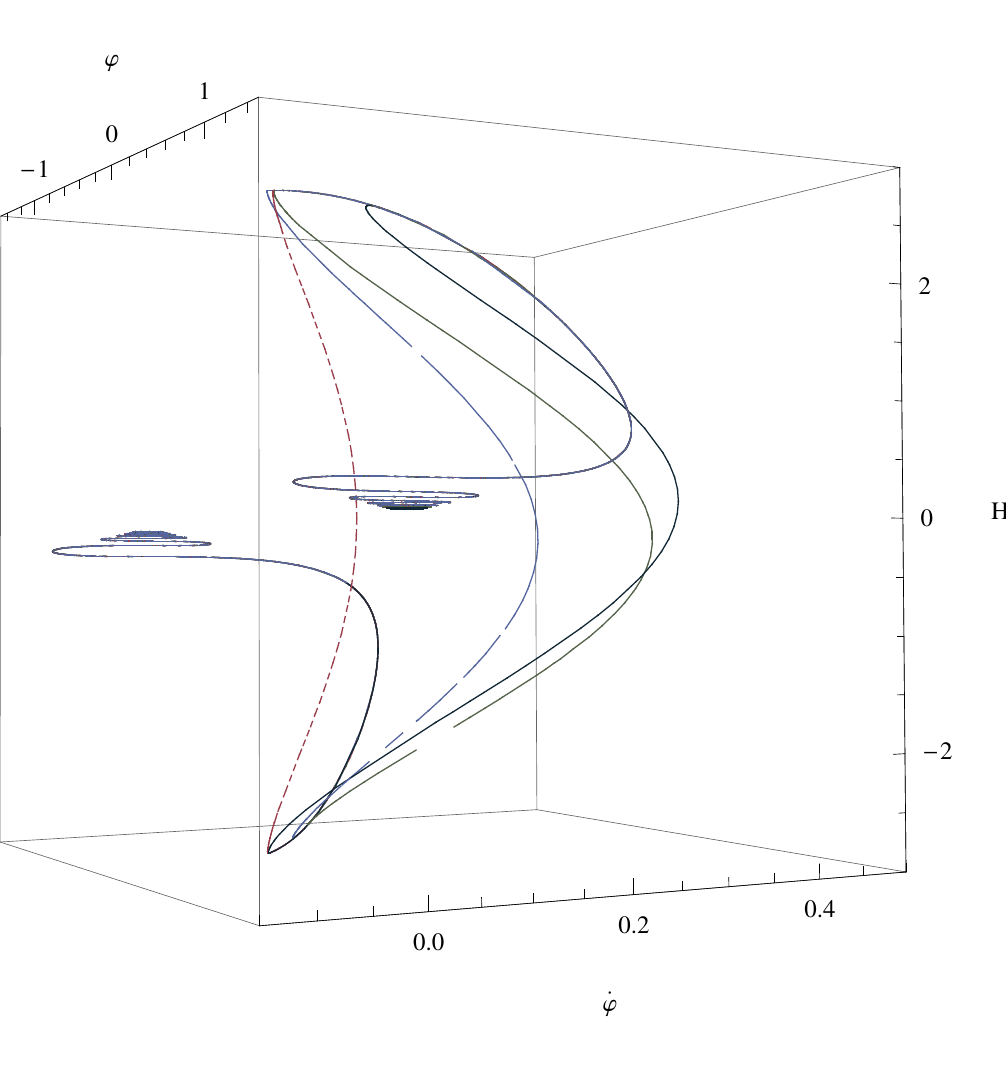}
\caption{Four trajectories joining a slow-roll contracting phase to a slow-roll expanding phase in the full three-dimensional phase space.  The sets of parameters used are the same as in previous figures.  These trajectories are well understood from combining the 2D trajectories of Fig.~\ref{PhaseSpace1} and Fig.~\ref{PhaseSpace2}. For a scalar field $\varphi$ at the bottom of its potential in a contracting phase, one has $H< 0$, $\varphi \simeq -2$ and $\dot{\varphi}\ne 0$.  Since $H<0$, if disturbed away from its minimum by a fluctuation, $\varphi$ will be driven farther away from its vacuum expectation value (see the bottom left panel of Fig.~\ref{PhaseSpace2} and also Fig.~\ref{traj2}).  For appropriately (fine) tuned initial conditions, the field will be driven towards $\varphi=0$ and $\dot{\varphi}\ll 1$ such that (see Fig.~\ref{PhaseSpace1}, going from left to right and also Fig.~\ref{traj1}) $H$ will grow more negative until $\varphi \simeq 0$, $\dot{\varphi} \ll 1$, where a slow-roll contracting phase (with $\varphi\simeq 0$, $\dot{\varphi}\ll 1$ and $H\simeq -\sqrt{V_0/3} \simeq 3$) will ensue.  The trajectory will then almost necessarily follow a bouncing trajectory and reach a slow-roll inflationary phase followed by a spiral down in the other minimum of $V\left( \varphi \right)$ (\ie it will follow an evolution identical to but reversed w.r.t. the contracting evolution just described). Note that, although it is certainly a possibility, here we do not consider the case in which the field turns around because as we will see, the perturbation equations are singular for $\dot{\varphi}=0$.}
\label{3D}
\end{figure}

Turning to the full $\left(\varphi,\dot{\varphi},\mathrm{H}\right)$ volume (see Fig.~\ref{3D} for an illustration of what a trajectory might look like), we may now determine both the location of the critical points and their stability.  The full differential system reads
\ba
\frac{\mathrm{d}\varphi}{\mathrm{d}t} & = & \dot{\varphi},\label{DiffSystem1}\\
\frac{\mathrm{d}^2\varphi}{\mathrm{d}t^2} & = & -3 H \dot{\varphi}+\mu^2 \varphi-\frac{\lambda}{6}\varphi^3,\label{DiffSystem2}\\
\frac{\mathrm{d}H}{\mathrm{d}t} & = & \frac{1}{3}\left(V_0-\frac{\mu^2}{2}\varphi^2+\frac{\lambda}{4!}\varphi^4-\dot{\varphi}^2\right)-H^2.\label{DiffSystem3}
\ea
The critical (fixed) points are determined by setting Eqs.~(\ref{DiffSystem1}) to (\ref{DiffSystem3}) to 0 and solving for $\varphi$, $\dot \varphi$, and $H$.  One finds
 \be
\begin{array}{cll}
\displaystyle \varphi=0, & \, \displaystyle \dot{\varphi}=0, & \, \displaystyle H=\pm\sqrt{\frac{ V_0}{3}},\\ \\
\displaystyle \varphi=\pm\sqrt{\frac{6 \mu^2}{\lambda}},& \, \displaystyle \dot{\varphi}=0, & \, \displaystyle H=\pm\sqrt{\frac{ \left(2V_0 \lambda-3\mu^4\right)}{6 \lambda}},\\ \\
\displaystyle \varphi=\pm\sqrt{\frac{6 \mu^2}{\lambda}},& \, \displaystyle \dot{\varphi}=0, & \, \displaystyle H=\mp\sqrt{\frac{ \left(2V_0 \lambda-3\mu^4\right)}{6 \lambda}}\nonumber.
\end{array}
\ee
Linearizing Eqs.~(\ref{DiffSystem1}) to (\ref{DiffSystem3}) around these six critical points and determining the corresponding set of eigenvalues provides a measure of stability.  We first find that $\left(0,0,\sqrt{  V_0/3}\right)$ is a saddle point SP1 with two negative eigenvalues $\sigma_1^\mathrm{(SP1)}$, $\sigma_2^\mathrm{(SP1)}$ and one positive eigenvalue $\sigma_3^\mathrm{(SP1)}$, given respectively by
\ba
\sigma_1^\mathrm{(SP1)} & = & -2\sqrt{\frac{  V_0}{3}},\\
\sigma_{2,3}^\mathrm{(SP1)} & = & \frac{1}{2}\left(-\sqrt{3  V_0}\mp
\sqrt{3  V_0+4\mu^2}\right).
\ea
Similarly, $\left(0,0,-\sqrt{  V_0/3}\right)$ is another saddle point SP2, this time with two positive eigenvalues  $\sigma_1^\mathrm{(SP2)}$, $\sigma_2^\mathrm{(SP2)}$ and one negative eigenvalue $\sigma_3^\mathrm{(SP2)}$, namely
\ba
\sigma_1^\mathrm{(SP2)} & = & 2\sqrt{\frac{  V_0}{3}},\\
\sigma_{2,3}^\mathrm{(SP2)} & = & \displaystyle \frac{1}{2}\left(\sqrt{3  V_0}\pm
\sqrt{3  V_0+4\mu^2}\right).
\ea
The four remaining critical points have eigenvalues
\ba
\sigma_1 & = & -2 H_0,\\
\sigma_{2,3} & = & \frac{1}{2}\left(-3H_0\pm\sqrt{9H_0^2-8\mu^2}\right),
\ea
where we have set $H_0=\sqrt{ \left(V_0/3-\mu^4/2\lambda\right)}$. One sees that  $9H_0^2-8\mu^2>0$ follows if $\lambda>9  \mu^4/\left[2\left(3   V_0-8\mu^2\right)\right]$, so that $\sigma_{2,3}<0$ (resp. $>0$) when $H_0>0$ (resp. $<0$), \ie they are either an attractor or a repulsor. 

On the other hand, $9H_0^2-8\mu^2<0$ implies that $\sigma_{2,3}$ are complex valued.  In this regime, they are either asymptotically stable (ASpP) or plainly unstable spiral points. The special situation $\lambda=3\mu^4/2V_0$ corresponds to $V\left(\varphi^+\right)=V\left(\varphi^-\right)=0$, \ie a vanishing cosmological constant $\Lambda=0$; the four distinct critical points then reduce to two simply stable spiral points with $H=0$.  In the following, largely for the sake of simplicity and since most of the physics we investigate takes place at the top of the potential, we set $\Lambda=0$.

As already suggested by Figs \ref{PhaseSpace1} and \ref{PhaseSpace2}, bouncing trajectories necessarily experience both a slow-roll inflationary phase, in the neighbourhood of the saddle point SP1 and a slow-roll contracting phase in the neighbourhood of the saddle point SP2. During inflation, the Hubble parameter reaches its maximum value $\sqrt{  V_0/3}$ ($\simeq 3$ in the figures) and then decreases along a ridge while $\varphi$ and $\dot{\varphi}$ will spiral into one of the ASpP.  Evolving the system backward in time, one finds that after crossing the $H=0$ plane the universe undergoes a slow-roll contracting phase with $H=-\sqrt{  V_0/3}$ ($\simeq -3$ in the figures) followed by a spiral towards one of the unstable spiral points.  The trajectory of Fig.~\ref{3D} is easily understood from combining trajectories in Figs \ref{PhaseSpace1} and \ref{PhaseSpace2}.  Indeed, begining with appropriately fine tuned initial conditions at $V_{\mathrm{min}}$ with $\varphi\simeq -2$, $\dot{\varphi}\ne 0$ and $H<0$, the system can naturally follow a trajectory that leads it towards $\varphi=0$ and $\dot{\varphi}=0$ (going from left to right in Fig.~\ref{PhaseSpace1}), $H$ will therefore become increasingly negative. This corresponds to the spiraling motion and the evolution along a ridge in Fig.~\ref{3D}.  This evolution drives the system to SP2, where slow-roll contraction occurs.  The system then naturally undergoes a bounce, reaches SP1, evolves out along a ridge (top right panel of Fig.~\ref{PhaseSpace2}), as $H$ decreases towards smaller positive values.  The spiralling into the ASpP then ensues.

At this stage of the analysis, the question of initial background conditions naturally arises.  Up to now, we have identified a finite phase space volume in which all trajectories bounce, two saddle points and four spiral points.  Working in this very same volume, one may further define the $\dot{H}=0$ surface on which lies a closed $\ddot{H}=0$ curve.  One can then identify a region on the portion of the surface bounded by the closed $\ddot{H}=0$ curve through which the dynamical system can escape the volume within and lead to a singular universe.  One finds that there is a sizeable region on the $\dot{H}=0$ surface for which $\ddot{H}<0$.  Combining volume and surface information, one concludes that initial conditions can be taken safely in the bulk only, or near the surface but in the region for which $\ddot{H}>0$.  The critical points $\left(0,0,\pm\sqrt{  V_0/3}\right)$ lie on both the $\dot{H}=0$ surface and the $\ddot{H}=0$ curve; so do $\left(\pm\sqrt{4V_0/\mu^2},0,0\right)$.  Given that the latter are spiral points, it turns out to be extremely difficult (at least numerically) to choose initial conditions leading to trajectories inside the favored volume: initial conditions taken in the neighborhood of these critical points but only very slightly away have a high chance of leading to trajectories that escape the volume and eventually become singular.

Let us now comment further on the required amount of fine-tuning necessary for the bounce to take place. Among the set $V_\mathrm{total}$ of all possible solutions, the subset that bounce, $V_\mathrm{b} \subset V_\mathrm{total}$ say, is either of dimension $D_\mathrm{b}=3$ or less ($D_\mathrm{b}<3$) thus leading to a fractal structure (see Refs.~\cite{H84,P84,CS98} for such considerations). In the latter case, the set of acceptable solutions is of zero measure, so one could argue, in the absence of a specific mechanism that would impose precisely those, that a bouncing phase could not have taken place. More work is needed to clarify this point. However, in the (more plausible?) case $D_\mathrm{b}=3$, the initial conditions needed to initiate a bouncing trajectory can be more easily driven to the allowed region (see however Ref.~\cite{Starobinsky78}). Not knowing what happens before the phases we describe here, it is essentially impossible to conclude on the amount of theoretical fine-tuning: the situation is akin to demanding vanishing spatial curvature $\Ka=0$, \ie a point (measure zero) on the line of possibilities or, invoking inflation, a small but finite region leading to an observationally small spatial curvature.

One point can however be discussed explicitly, and it concerns the required level of numerical precision needed to reach a bouncing solution starting from a large contracting universe. Let us assume that we begin the calculation at some time $t_\mathrm{ini}$ say, where we want to impose initial conditions. Writing the Friedmann constraint equation at $t_\mathrm{ini}$ yields
\be
a_\mathrm{ini} \equiv \ex^{N_\mathrm{ini}} = 
\left\{
\frac{\Ka}{\left[ \frac12 \dot{\varphi}_\mathrm{ini}^2 + V\left(\varphi_\mathrm{ini} \right)
\right] - H_\mathrm{ini}^2}
\right\}^{1/2},
\ee
with $N_\mathrm{ini}$ the number of e-folds we start with. It is clear that the denominator is of
order $\ex^{N_\mathrm{ini}}$. In other words, the larger the universe prior to contraction, the more numerical accuracy is necessary at $t_\mathrm{ini}$ to satisfy the constraint equation. And this is even before one asks the question of the Lyapunov stability of the dynamical system.  Again, all these considerations do strongly depend on the prior mechanism of precooling~\cite{precool:2008}, which might in fact have the ability to naturally lead to the nonsingular region.

\section{Scalar Perturbations}\label{sec:perturbations}

\begin{figure}[t]
\begin{center}
\includegraphics[width=9cm]{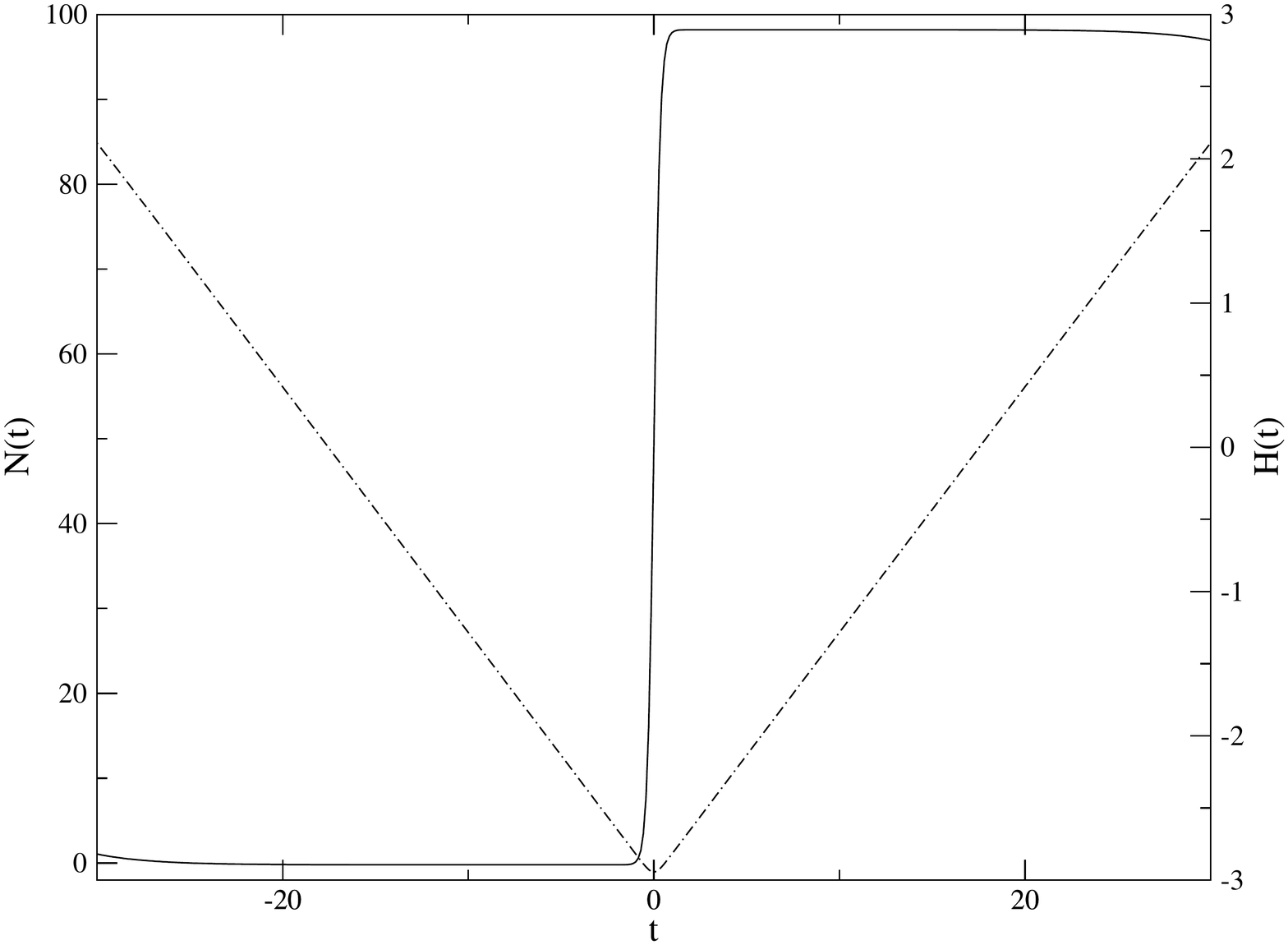}
\end{center}
\caption{Number of e-folds, $\log a$, (dashed-dotted)  and Hubble parameter $H$ (full) as functions of time, for a symmetric bounce, with the Mexican-hat potential $V\left( \varphi \right)$ of Eq.~(\ref{V2}), with parameters $V_0=1$, $\mu=1$ and $\lambda=3   \mu^4/2 V_0 $. In order to obtain this solution, initial conditions were taken at the bounce, with $\dot{\varphi_0}=1/10$. Note that these values were chosen such that the Hubble parameter remains almost constant in both the contracting and expanding phases for $\approx 70$ e-folds.}
\label{background}
\end{figure}

We now consider a typical bouncing solution, such as the one depicted on Fig.~\ref{background} where the solid and dashed lines represent the time evolution of $H$ and $a$ respectively, from a slow-roll contracting phase across a bounce to a slow-roll inflationary phase.

Let us now move on to metric perturbations, and consider the gauge invariant gravitational
potential $\Phi$ in the presence of density perturbations $\delta \varphi$.  Working in
conformal time $\eta$ for convenience, and in longitudinal gauge~\cite{KS84,MFB92}, the
scalar part of the perturbed metric reads
\be
\mathrm{d}s^2 = a^2\left(\eta \right) \left[-\left(1+2\Phi\right)\dd\eta^2 + \left(1-2\Phi \right)
\gamma_{ij} \mathrm{d}x^i\mathrm{d}x^j\right].
\ee
where $\gamma_{ij}$ is the background metric of the spatial sections.

Einstein's equations, to first order, imply
\be
\displaystyle \Phi''+2\left(\mathcal{H}-\frac{\varphi''}{\varphi'}\right) \Phi' + \left[k^2-4\mathcal{K}+2\left(\mathcal{H}'- \mathcal{H}\frac{\varphi''}{\varphi'}\right)\right]\Phi=0,
\label{Bardeen}
\ee
where $k=\sqrt{n(n+2)}$ is the comoving wavenumber (the eigenvalue of the Laplace-Beltrami operator),
while $\Hu$ denotes the conformal time counterpart of $H$, namely $\Hu\equiv a'/a = aH$.

Using the generalized form of the Mukhanov-Sasaki variable $Q$~\cite{MC811,MC812,Sasaki83}
\be
Q=\delta\varphi+\left(\frac{ \varphi'^2-2\mathcal{K}}{ \mathcal{H}^2\varphi'}\right)\Phi,
\label{MukhanovSasaki}
\ee
and further defining $\tilde{Q}=aQ$, Eq.~(\ref{Bardeen}) can be written in the form
\be
\tilde{Q}''+\left(\frac{2\chi'}{\chi}-\frac{z'}{z}\right)\tilde{Q}'+
\left(\frac{z}{\chi^2}+\frac{\chi''}{\chi}-\frac{z'\chi'}{z\chi}\right)\tilde{Q}=0,
\label{QEquationK}
\ee
where $$\chi\equiv\frac{\mathcal{H}}{a\varphi'},$$ and $$z\equiv\chi\left[k^2+\Ka \left( 6+
\frac{2\varphi''}{\mathcal{H}\varphi'} \right)\right].$$
Note, incidentally, that Eq.~(\ref{QEquationK}) reduces to the usual form
\be
\tilde{Q}''+\left[k^2+a^2V_{,\varphi\varphi}+\frac{2\mathcal{H}''}{\mathcal{H}}-2\left(\frac{\mathcal{H}'}{\mathcal{H}}\right)^2+\mathcal{H}'-5\mathcal{H}^2\right]\tilde{Q}=0\label{Qflat}
\ee
when the spatial sections are flat.

It is obvious from Eqs.~(\ref{Bardeen}) and (\ref{Qflat}) that the equation of motion for
$\Phi$ is well behaved through the bounce, even though $\mathcal{H}$ goes through zero,
while that for $\tilde{Q}$ is well behaved far from the bounce in the so-called reheating
(precooling) phase, that follows (precedes) the inflationary (contracting) phase
in which $\varphi'$ oscillates around zero. Eq.~(\ref{QEquationK}), however, appears well
behaved in neither one of these two regimes.

\begin{figure}
\begin{center}
\includegraphics[width=9cm]{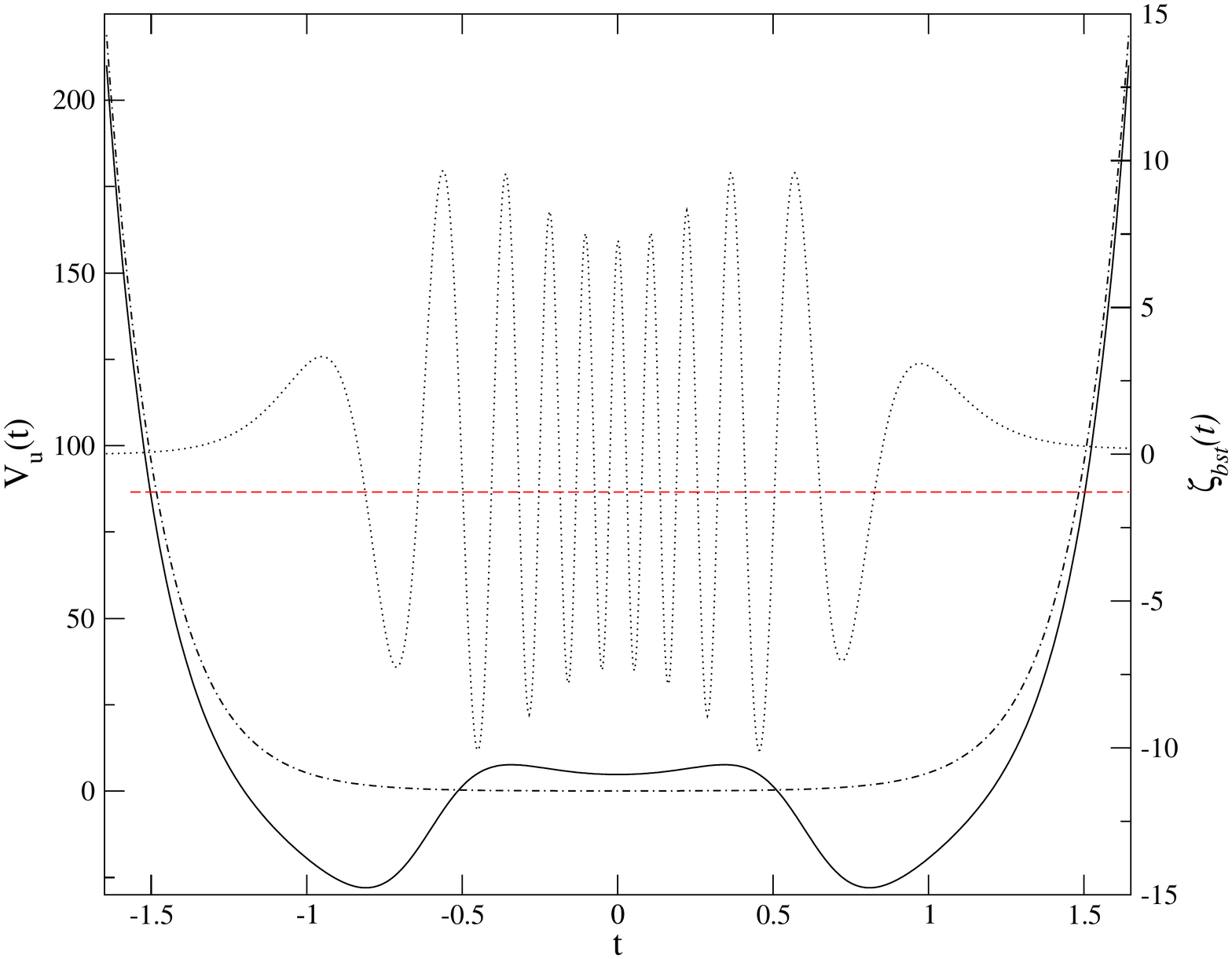}
\end{center}
\caption{The potential $U$ (solid line), and its approximation (dashed-dotted) in the slow-roll region, together with $\zeta_\mathrm{BST}$ (dotted) as functions of time, obtained from the evolution of $u$ using (\ref{u}) in the background of Fig.~\ref{background}, with $n=20$. We should emphasize that, even though this is not immediately obvious from the graph, $\zeta_\mathrm{BST}$ is not conserved across the bounce.}
\label{perturbations}
\end{figure}

In fact, Eq.~(\ref{Qflat}) can be made more compact by taking
$z_0\equiv a\varphi'/\mathcal{H}$,
in which case one obtains the well-known form of the equation for the perturbations as that of a parametric
oscillator,
\be
\tilde{Q}''+\left(k^2-\frac{z_0''}{z_0}\right)\tilde{Q}=0.
\label{Qsimple}
\ee
This well-known form is indeed simpler than (\ref{Qflat}), but the potential term is of course still singular when $\mathcal{H}=0$ and Eq.~\ref{Qsimple} is therefore not useful through the bounce. The inverse set of remarks can be made of the often used variable $u$, which is related to $\Phi$ by
\be
\Phi=\frac{3 \Hu}{2 a^2\theta}u,
\label{Phi2u}
\ee
where
\be
\theta=\frac{1}{a}\left(\frac{\rho_{\varphi}}{\rho_{\varphi}+p_{\varphi}}\right)^{1/2}
\left(1-\frac{3\mathcal{K}}{ \rho_{\varphi}a^2}\right)^{1/2},
\ee
leading to an equation of motion for $u$ which reads
\be
u''+\left[ k^2-\frac{\theta''}{\theta}-3\mathcal{K}\left(1-\cs\right)\right]u=0\label{u},
\ee
where the potential term is $U(\eta)=\theta''/\theta+3\mathcal{K}\left(1-\cs \right)$, and given explicitely by 
\be
U(\eta)=\mathcal{H}^2+2\left(\frac{\varphi''}{\varphi'}\right)^2-
\frac{\varphi'''}{\varphi'}-\mathcal{H}'+4 \mathcal{K}.
\ee
 Note that this potential grows as $a^2$ away from the bounce and is not well-defined in the oscillatory preheating (cooling) phase that follows (precedes) the bounce.  The virtue of using the variable $u$ in fact lies in the fact that Eq.~(\ref{u}) is well behaved at the bounce, as opposed to Eq.~(\ref{Qsimple}).  The form of Eq.~(\ref{u}) is therefore particularly well suited to understand the effect of the potential term $U(\eta)$ in the neighborhood of the bounce.   Indeed, substituting Eqs.~(\ref{eta0}), (\ref{delta}) and (\ref{xi}) into the asymmetric version of Eq.~(48) of Ref.~\cite{MP03}, the amplitude of the potential at the time of the bounce reads
\be
\label{Vu0}
U_\mathrm{b}= \left(5-\Upsilon\right)+\frac{V_\mathrm{b}''}{V_\mathrm{b}}\left(3-\Upsilon\right)
 +\left(\frac{V_\mathrm{b}'}{V_\mathrm{b}}\right)^2\frac{\left(\Upsilon-3\right)^2}{\Upsilon},
\ee
so that taking the limit $\Upsilon \to 0$ leads to $\dot{\varphi} \to 0$ and $\delta \to 0$ and approaches a symmetric de Sitter bounce.  Note that we have switched notation: what was labelled using the subscript `$_0$' in \S~\ref{sec:background} and denoted quantities evaluated at the bounce now bares the subscript `$_{\mathrm{b}}$' in order to distinguish these quantities from $V_0$ which now appears explicitly in the following expressions as the constant term of the potential $V \left( \varphi \right) \equiv V(\varphi=0)$.  Using Eq.~(\ref{V2}) one has
\be
\frac{V_\mathrm{b}'}{V_\mathrm{b}}=\left[ \frac{\varphi}{4}+\frac{3\left(\mu^2\varphi^2-4V_0\right)}{12\mu^2\varphi-2\lambda \varphi^3}\right]^{-1},
\ee
and
\be
\frac{V_\mathrm{b}''}{V_\mathrm{b}}=\left(\frac{\varphi^2}{12}-\frac{5\mu^2}{6\lambda}-
\frac{6V_0-5\mu^2}{6\lambda\mu^2-3\lambda^2\varphi^2}\right)^{-1},
\ee
such that if the scalar field goes to either one of the limits
\be
\varphi \to \pm \sqrt{\frac{6\mu^2}{\lambda}\pm\frac{2\left(9\mu^4-6V_0\lambda
\right)^{1/2}}{\lambda}},
\label{PhiMax}
\ee
the potential exhibits a large central peak.  However, for the theory to make any sense at all, the potential should be positive-definite. This is ensured provided one demands $V_{\mathrm{min}} \ge 0$, where $V_{\mathrm{min}} = V_0 -3\mu^4/(2\lambda)$ is the minimum value of $V$, attained for $\varphi\to \sqrt{6\mu^2/\lambda}$. This means that $\lambda \ge 3\mu^4/(2V_0)$. In Eq.~(\ref{PhiMax}) however, the argument of the square root requires the opposite condition $\lambda \le 3\mu^4/(2V_0)$ to hold in order to keep $\varphi$ real. The abovementioned limit can therefore not be  approached without violating one or the other condition: the quantities $V_{\mathrm{b}}'/V_{\mathrm{b}}$ and $V_{\mathrm{b}}''/V_{\mathrm{b}}$ must therefore remain small. The potential $U_{\mathrm{b}}$, having no large central peak, therefore does not affect the perturbations much as they pass through the bounce itself.  The question of the wings (located at $\sim \pm 0.75$ in Fig.~\ref{perturbations}), and in particular whether there is any way of showing they remain small, is left unanswered.

It should be noted that the variable $\tilde{Q}$ is equivalent to the variable often denoted by $v$ in the literature (see, \eg Ref.~\cite{MFB92}). It arises naturally when expanding the Einstein action up to second order in the perturbations, giving the action of a scalar field in flat space having a varying mass (\ie a pump term) caused by gravity. It allows an easy definition of the initial quantum state.  Assuming adiabatic vacuum in the low-curvature
regime for which $k^2\gg z''/z$ in Eq.~(\ref{Qsimple}), one sets
\be
\tilde{Q}_\mathrm{init} \propto\frac{1}{\sqrt{2k}}\ex^{-ik\eta}.
\label{InitialState}
\ee
We have already seen in Section \ref{sec:phase} that the initial condition for the background is quite difficult to achieve in a natural way. We now encounter another weakness of curvature-based bouncing models such as this one, namely that there does not seem to be a generic way of setting up appropriate initial conditions for the perturbations. This is due to the fact, revealed by Eq.~(\ref{Qflat}), that, because the potential $U$ never drops to negligibly small values far from the bounce, the condition $k^2\gg z''/z$ is never satisfied because $a^2V_{,\varphi\varphi}$ dominates.

What, then, should we choose as initial conditions for the perturbations?

In slow-roll inflation, $\Phi=\mathrm{const.}$ in the regime $k \ll a H$, such that one has $\dot{\Phi}=0$.  Fixing an initial condition in this regime seems hardly motivated indeed, as it would be equivalent to assuming an initial spectrum, contrary to the basic original idea of computing this spectrum out of a natural prescription.  However, the model we are considering here cannot be made complete, in the sense that a contracting epoch should be connected to the bounce through a yet undescribed precooling phase~\cite{AP07}. Let us then press on and assume the usual form
\be
\Phi \propto C(k) \left[1-\frac{H}{a}\int_{t_\star}^t a\left(\tilde{t} \right)
\mathrm{d}\tilde{t}\right],
\label{AsymptoticBehaviour}
\ee
in which the lower bound $t_\star$ on the integration is unknown and we have parameterized our ignorance through the unknown overall multiplicative function of the wavenumber $k$, $C(k)$.

As first recognized in~\cite{WMLL00a}, Eq.~(\ref{Bardeen}) admits a first integral given by
\be
\zeta_\mathrm{BST}=-2\frac{H^2}{  \dot{\varphi}^2}\left[\frac{\dot{\Phi}}{H}+
\left(1-\frac{\Ka}{a^2H^2}+\frac{1}{3}\frac{k^2}{a^2H^2}\right)\Phi \right]-\Phi,
\ee
such that~\cite{MS98}
\be
\dot{\zeta}_\mathrm{BST}\propto\frac{k^2}{3a^2H^2}\left( \dot{\Phi}+H \Phi \right)
+\frac{1}{2}\frac{\tau\delta S}{H},
\label{zetadot}
\ee
where the nonadiabatic pressure perturbation is given in terms of the
entropy perturbation $\delta S$ through
\be
\frac{1}{2}a^2\tau\delta S \equiv \delta p_{\varphi}-\cs\delta \rho_{\varphi} 
= \left(1-\cs\right)\left(3\Ka-k^2\right)\Phi,
\ee
with $\cs$ defined by the adiabatic variation of $\rho_{\varphi}$ and $p_{\varphi}$ as
\be
\cs \equiv \left(\frac{\delta p_{\varphi}}{\delta \rho_{\varphi}}\right)_S
= -\frac{1}{3}\left(3+2\frac{\ddot{\varphi}}{H\dot{\varphi}}\right)
\label{cs2}
\ee
and is the square of the sound velocity, \ie the velocity at which perturbations propagate.

For a single scalar field in flat space, $\delta S\propto k^2 \Phi$, as a result of which the entropy perturbation is essentially negligible with respect to the adiabatic ones, and $\dot{\zeta}$ vanishes as long as $k \ll a$. The curvature perturbation $\zeta$ thus remains constant on super-Hubble scales. In the inflationary scenario, one uses the variable $v$ and the dynamical equation for the perturbations can be solved directly in terms of Bessel functions of the first kind of order $\alpha$ where $\alpha$ is a constant that can be expressed in terms of the slow-roll parameters $\varepsilon$ and $\delta$, namely $\alpha \sim -3/2-2\varepsilon+\delta$.  Using the asymptotic behaviour of the Bessel function and performing two consecutive matchings with the vacuum solution of Eq.~(\ref{InitialState}) for $k \eta\rightarrow -\infty$ on the one hand, and with a solution of the type (\ref{AsymptoticBehaviour}) for $k \eta\rightarrow 0$ on the other, one obtains
\be
{\mathcal P}_{\zeta}=k^3 |\zeta|^2 \sim k^3 A^2(k) k^{2\alpha+1},
\ee
where $A(k)$ is the power law $k$ dependence of the initial state. One sees that
the standard slow-roll spectrum $k^{-4 \varepsilon+2\delta}$ stems directly from the
Minkowski vacuum, in which $A(k) \sim k^{-1/2}$. 

In the bounce scenario, the situation is slightly different, as, in particular, no specific prescription
for the initial state exists.  Given an appropriate potential $U$, one may however similarly perform several consecutive matchings.  We have already shown that the potential for $u$ at the bounce is negligible.  This means that the solution in the neighborhood of the bounce can be taken as a linear combination of oscillatory functions, with unknown $k$-dependent coefficients that depend on the initial state. In the slow-roll regions on either sides of the bounce, where curvature can be neglected, the solution is a linear combination of Bessel functions while very far from the bounce,  where $k \ll aH$, it is given by Eq.~(\ref{AsymptoticBehaviour}).  In practice, it turns out to be sufficient to perform only two matchings. The first joins the initial conditions (for which a reasonable prescription has yet to be given) to a solution in terms of Bessel functions, while the second joins Bessel functions solutions having arguments $k|\eta-\eta_-|$ (respectively $k|\eta-\eta_+|$) and order $\alpha_-$ (resp. $\alpha_+$) where $\eta_\pm$ and $\alpha_\pm$ are defined using the approximate form of the potential far from the bounce. In the slow-roll regions, for which the curvature term is unimportant, being suppressed by a factor $a^2$, the potential for $u$ can be expressed as 
\be
U=a^2 H^2\left(2 \varepsilon -\delta+2 \varepsilon^2+\delta^2-3 \varepsilon \delta-\xi \right)
\ee
where $\varepsilon$, $\delta$ and $\xi$ have their usual definitions in terms of the derivatives of
the potential w.r.t. the field~\cite{STG01,ST04}. In such a case, the
potential  term before  ($U_-$) and after ($U_+$) the bounce reads
\be
U_\pm  = \frac{\alpha_\pm}{\left(\eta-\eta_\pm\right)^2},
\ee
where $\eta_-$ and $\eta_+$ are the asymptotic limits to the left and right of the bounce
respectively.  In de Sitter $\eta_{\pm}=\pm \pi/2$.  The indices $\alpha_\pm$ are obtained
from the slow-roll parameters $\varepsilon_{\pm}$, $\delta_{\pm}$ and $\xi_{\pm}$ as
\be
\displaystyle 4 \alpha_\pm^2 = \displaystyle 1+4 \left(2 \varepsilon_\pm -\delta_\pm
+2 \varepsilon_\pm^2+\delta_\pm^2-3 \varepsilon_\pm \delta_\pm-\xi_\pm 
\right),\ee
which, for small values of  $\varepsilon_{\pm}$, $\delta_{\pm}$ and $\xi_{\pm}$ gives
\be
\displaystyle \alpha_\pm \simeq  \frac12 + \left(2 \varepsilon_\pm -\delta_\pm+2 \varepsilon_\pm^2+
\delta_\pm^2-3 \varepsilon_\pm \delta_\pm-\xi_\pm\right).
\ee
For a symmetric potential $V(\varphi)$, provided sustained slow-rolling contracting and
expanding phases are achieved, $\alpha_-$ and $\alpha_+$ are equal, because Fig.~\ref{3D}
is symmetric w.r.t. the $(\varphi$, $\dot{\varphi})$ plane. This is clearly true in particular in
the case of a symmetric bounce, but most likely also even in most asymmetric bounces.

In order to set up sensible initial conditions away from the bounce, \ie where $k^2 \ll U $, one
has, using Eqs.~(\ref{Phi2u}) and (\ref{AsymptoticBehaviour}),
\be
u\propto \frac{C(k)}{\varphi'} \left[1-\frac{\mathcal{H}}{a^2\left(\eta \right)}\int_{\eta_\star}^{\eta}a^2\left(\tilde{\eta} \right) \mathrm{d}\tilde{\eta} \right].
\ee
Here, $C(k)\propto k^{\gamma}$, and both $\gamma$ and $\eta_\star$ are constants that depend on a preceding precooling phase.  In the de Sitter limit, one obtains the approximate initial condition
\be
u_\mathrm{i} \propto C(k) \frac{\mathcal{H}_i}{\varphi_i' a_i}.
\ee
Note also that far from the bounce,
\be
\zeta_\mathrm{BST}=\zeta=-\frac{\mathcal{H}}{a \varphi'} 
\left(u'+\frac{\varphi''}{\varphi'}u\right)-\frac{1}{2} \frac{\varphi'}{a}u,
\ee
so that $\zeta\propto C(k)$, a mere function of the wavenumber $k$, \ie a constant in time, the $k$ dependence of which 
is the same as that for $u$. The matching is performed from a slow-rolling contracting
phase through the bounce into an expanding slow-rolling phase, \ie from 
\begin{widetext}
\be
u\left(\eta\right)=\sqrt{k \left(\eta -\eta_-\right)} \left\{A_1 J_{\alpha_-}
\left[(k \left(\eta -\eta_-\right) \right] +A_2 J_{-\alpha_-}
\left[k \left(\eta -\eta_-\right)\right] \right\},
\ee
where 
\be
A_1=\frac{2^{-2+\alpha_-} \pi \left[k\left(\eta_\star-\eta_-\right)\right]^{-1/2-\alpha_-} 
\left[u_i \left(2 \alpha_- -1\right)+2 u'_i \left(\eta_\star-\eta_-\right)\right] \mathrm{csc}
\left(\pi \alpha_-\right)}{\Gamma\left(1-\alpha_- \right)}
\ee
and
\be
A_2=2^{-2-\alpha_-} \left[k\left(\eta_\star-\eta_-\right)\right]^{-1/2+\alpha_-} 
\left[u_i \left(2 \alpha_-+1\right)-2 u'_i \left(\eta_\star-\eta_-\right)\right]
\Gamma\left(-\alpha_-\right),
\ee
to
\be
u\left(\eta\right)=\sqrt{k \left(\eta -\eta_+\right)}
\left[B_1\left(k\right) J_{\alpha_+} (k \left(\eta -\eta_+\right))+B_2
\left(k\right) J_{-\alpha_+} (k \left(\eta -\eta_+\right)\right]
\ee
where
\be
B_1\left(k\right)=-\mathrm{csc}\left(\pi \alpha_+ \right)
\left\{ A_1 \mathrm{cos}\left[k\left(\eta_--\eta_+\right)+
\frac{\pi}{2}\left(\alpha_--\alpha_+ \right)\right]+A_2 \cos
\left[k\left(\eta_- -\eta_+\right)-
\frac{\pi}{2}\left(\alpha_-+\alpha_+\right)\right]\right\}
\ee
and
\be
B_2\left(k\right)=\mathrm{csc}\left(\pi \alpha_+ \right) 
\left\{ A_1 \cos \left[k\left(\eta_--\eta_+\right)+
\frac{\pi}{2}\left(\alpha_-+\alpha_+ \right)\right]+A_2 \cos
\left[k\left(\eta_- -\eta_+\right)-\frac{\pi}{2}
\left(\alpha_--\alpha_+\right)\right]\right\}.
\ee
The mixing matrix is thus given by
\begin{displaymath}
\left( \begin{array}{c}
B_1\\ \\
B_2
\end{array}
\right)
=\mathrm{csc}\left(\pi \alpha_+ \right)
\left( \begin{array}{ccc}
\mathrm{cos}\left[k\left(\eta_--\eta_+ \right)+\frac{\pi}{2}\left(\alpha_- -\alpha_+ \right)\right] &\null &
 \mathrm{cos}\left[k\left(\eta_--\eta_+ \right)-\frac{\pi}{2}\left(\alpha_- +\alpha_+ \right)\right]\\ & & \\
\mathrm{cos}\left[k\left(\eta_--\eta_+ \right)+\frac{\pi}{2}\left(\alpha_- +\alpha_+ \right)\right] & 
\null & \mathrm{cos}\left[k\left(\eta_--\eta_+ \right)-\frac{\pi}{2}\left(\alpha_- - \alpha_+ \right)\right]
\end{array}
\right)
\left( \begin{array}{c}
A_1\\ \\
A_2
\end{array}
\right).
\end{displaymath}
\end{widetext}
This result differs from the one obtained in~\cite{MP03} in two ways. First, the characteristic time over which the perturbations may be altered by the bounce: in~\cite{MP03}, it was extremely small, and the effect was mainly concentrated at the bounce itself \footnote{A $\delta$ function approximation for the potential was even suggested.}, whereas in the present case, taking into account the quasi-de Sitter contraction and expansion phases that take place before and after the bounce, the actual characteristic time scale turns out to be $\eta_--\eta_+$. The second crucial difference is that the bounce itself leaves the perturbations unchanged, since the potential is negligible during this stage of cosmological evolution. Our result however agrees with Ref.~\cite{MP03} in the sense of Eq.~(70) of the abovementionned reference, \ie in the immediate neighborhood the bounce, one gets the identity, together with some possible overall rescaling of the total amplitude.  We however expect the result presented here to hold more generically but we remind the reader that as already mentioned, the question of the amplitude of the wings of the potential is left unanswered.

Let us turn to the power spectrum. For $k(|\eta-\eta_+|)\ll1$, the first term is negligible, and the result can be approximated by the term in $B_2$. In addition, $A_2$ also turns out to be negligible.  One thus has
\be
\mathcal{P}_{\zeta} \propto k^{1+2 \gamma-2 \delta \alpha_--2 \delta \alpha_+}
\cos^2 \left[ k \left(\eta_- - \eta_+ \right)+\frac{\pi}{2} \left(\alpha_-+\alpha_+ \right) \right],
\ee
where $k^{\gamma}$ is the initial $k$ dependence of $u_i$ and $u'_i$, \ie that of $C(k)$.  The power-law coefficients $\delta\alpha_\pm$ are given by
\be
\delta \alpha_\pm = 2\varepsilon_\pm - \delta_\pm +2 \varepsilon_\pm^2+\delta_\pm^2-3\delta_\pm \varepsilon_\pm - \xi_\pm.
\ee
From the above, we find that as $\dot\varphi$ gets vanishingly small, then
$\pi-(\eta_--\eta_+) \rightarrow \pi^-$, and both $\delta \alpha_\pm\rightarrow 0$ as $\mu$ is decreased.
\begin{figure}
\begin{center}
\includegraphics[width=8.5cm]{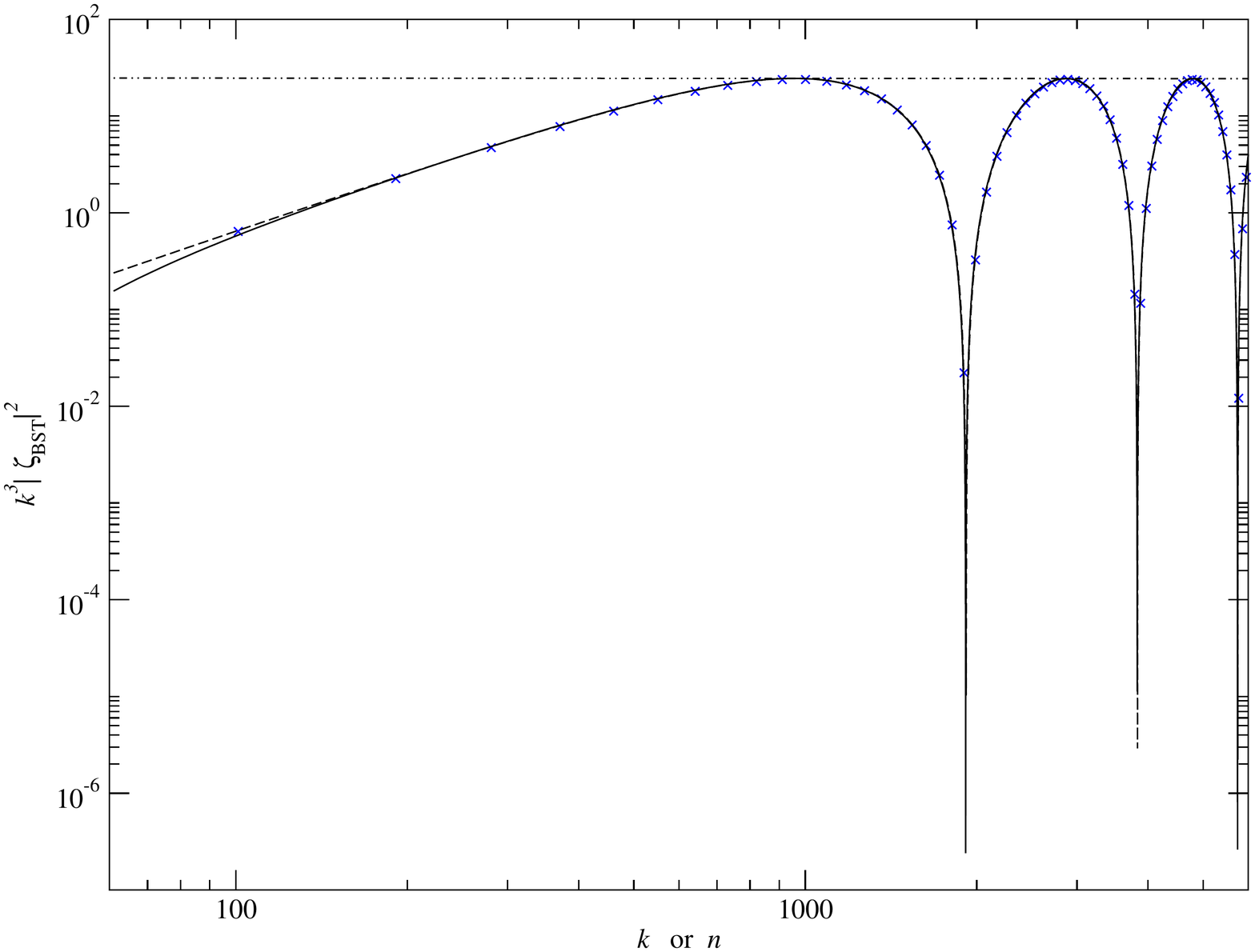}
\end{center}
\caption{Expected spectrum of the curvature perturbation $\zeta_\mathrm{BST}$. The theoretical prediction of Eq.~(\ref{powerspectrum}) is the solid line, while its approximation of Eq.~(\ref{powerspectrumapprox}) is the dashed line. The crosses are the corresponding numerical results obtained by evolving Eq.~(\ref{u}) for wavenumbers $n$ in the range $n\in [60-2000]$.  The power law term alone is shown in the dashed-dotted line.  Here, the parameters for the potential $V(\varphi)$ of Eq.~(\ref{V}) are $\mu=1/5$, $\dot{\varphi}=1/30$; these parameters imply $\delta \alpha_+=\delta \alpha_- \approx 10^{-3}$ and $\delta \eta \approx 10^{-3}$.}
\label{spectrum}
\end{figure}
In the special case of $\gamma=-1/2$ (in which case one starts out with an almost scale-invariant spectrum for $\mathcal{P}_{\zeta}$ far in the past prior to the bounce), the resulting power spectrum takes the form 
\ba
\mathcal{P}_{\zeta} & \propto & k^{-2 \delta \alpha_--2 \delta \alpha_+} \cos^2 \left[ k \left(\eta_- - \eta_+ \right)+\frac{\pi}{2} \left(\alpha_-+\alpha_+ \right) \right] \nonumber \\
& \propto & k^{-2 \delta \alpha_- -2 \delta \alpha_+} \sin^2 \left[ k \left(\pi+\delta \eta \right)+\frac{\pi}{2}\left(\delta \alpha_-+\delta \alpha_+\right) \right]\nonumber .\\
\label{powerspectrum}
\ea
where we have set, for notational simplicity,
\be
\delta \eta \equiv \pi-(\eta_--\eta_+).
\ee
Taking $k=\sqrt{n \left(n+2\right)}\approx n+1-1/2n$, one obtains the approximate form
\be
\mathcal{P}_{\zeta} \propto k^{-2\left( \delta \alpha_-+ \delta \alpha_+\right)} \sin^2 \left(n \delta \eta \right).
\label{powerspectrumapprox}
\ee
which we compare, on Fig.~\ref{spectrum}, with Eq.~(\ref{powerspectrum}) and with the actual full numerical solution. It is clearly a satisfying approximation as the points and curves are almost undistinguishable.  Not only does the bounce modify the power law behavior of $\mathcal{P}_{\zeta}$ but it also adds to it a multiplicative oscillatory term at leading order.  In first approximation, one further has
\be
\delta \eta \simeq \displaystyle \pi \left( \frac{1}{\sqrt{a_0^2 V_0-2\Ka}}-1 \right).
\ee
This highlights the fact that this bouncing scenario is fully characterized by the usual slow-roll parameters encoded in $\delta \alpha_{\pm}$ and by the scale factor at the bounce, $a_0$, that appears in the expression for $\delta \eta$.
The above oscillating spectrum can be rephrased, once the inflation epoch has taken place and ended, into a more generic form
\be
\mathcal{P}_\zeta = \mathcal{A} k^{\ns-1} \cos^2 \left( \omega \frac{k_\mathrm{ph}}{k_\star} +\varphi \right),
\label{cos2}
\ee
where now the wavenumber $k_\mathrm{ph}$ is written in units of inverse length, the nominal scale $k_\star$ can be fixed at the given scale of 100 Mpc and $\varphi$ is some phase. One is thus left with one arbitrary parameter, namely the oscillation frequency $\omega$. For very small frequencies, \ie $\omega\ll 1$, the cosine term remains constant and this is nothing but a simple power-law spectrum, and no new information can be derived. On the other hand, in the large frequency limit, $\omega\gg 1$, the resulting spectrum, once evolved in the radiation and matter dominated eras and expressed in terms of CMB multipoles, is not immediately ruled out, contrary to what might have been expected first hand.  Fig.~\ref{osc} shows that for the present best set of data from first year~\cite{Peiris:2003ff} or third year WMAP~\cite{Hinshaw+07}, a very reasonable fit can be obtained~\cite{Martin:2003sg,Martin:2004iv}. Clearly, with a primordial spectrum as the one exhibited on Fig.~\ref{spectrum}, one might have naively thought it impossible to recover a temperature perturbation spectrum such as shown on Fig.~\ref{osc}.
It turns out, however, that provided the frequency $\omega$ is sufficiently large the spherical Bessel smoothing functions used to derive the latter from the former~\cite{PPJPU} allow for a fit which is neither compellingly wrong, nor any better that the usual power-law spectrum.  Similarly, in the observable range of scales where only linear interactions play a role in the eras following inflation, and for which the transfer function from the primordial power spectrum is a simple function, oscillations in the density power spectrum can nonetheless be expected to be smoothed out by a convolution with a windowing function linked to survey size in observations.  Constraining the parameters of this model using large-scale galaxy surveys and CMB data will be discussed in future work.  In particular, relating the parameters $\omega$, $\ns$ and $\mathcal{A}$ in Eq.~(\ref{cos2}) to those of our spectrum (\ref{powerspectrumapprox}), and to those appearing in the potential (\ref{V2}), namely $V_0$, $\mu$, and $\lambda$ will be needed to go any further and place meaningful constraints on this particular type of model.

\begin{figure}[h]
\begin{center}
\includegraphics[width=8.5cm]{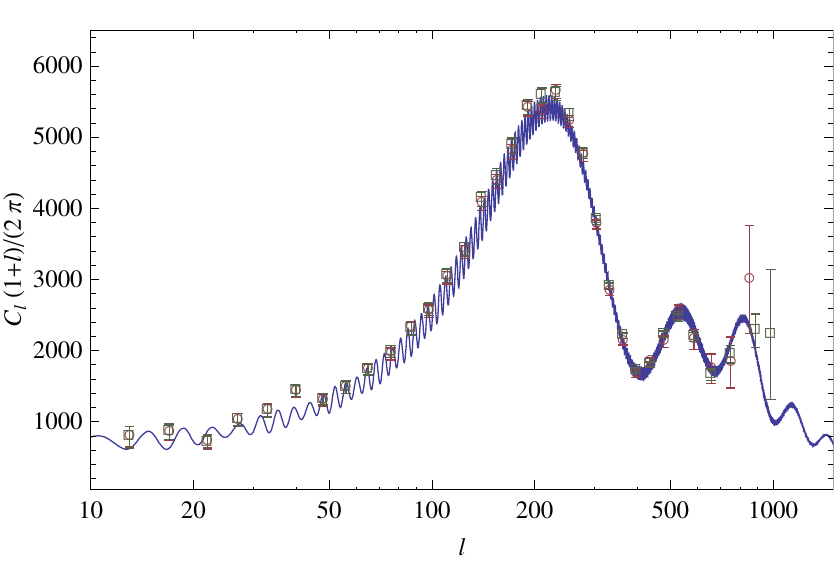}
\includegraphics[width=8.5cm]{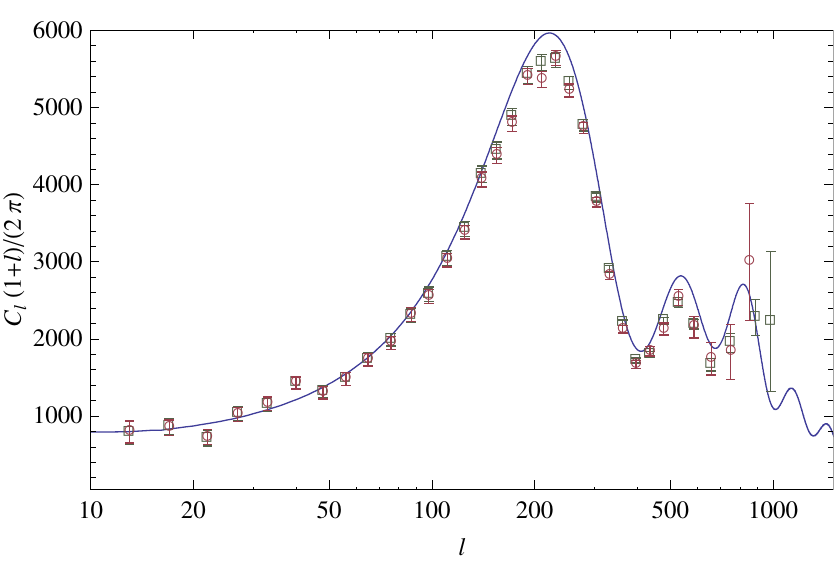}
\end{center}
\caption{Using the spectrum of Fig.~\ref{spectrum} is not obviously incompatible with the data: WMAP data (1 and 3 years, respectively, squares and circles) are superimposed on the predicted (full line) temperature fluctuations stemming from (\ref{cos2}) for fixed $k_{\star}$, $\omega=2000$ and $1000$ for the upper and lower plots respectively and $\varphi$ was set to zero for simplicity.  Both the amplitude and frequency of oscillations are affected by varying the parameters of the model. Figure courtesy of C. Ringeval~\cite{priv:CRFig}.}
\label{osc}
\end{figure}

%\newpage
\section{Conclusions}

Using the simplest possible theoretical framework in which a bouncing Friedmann-Lema\^{\i}tre-Robertson-Walker scale factor is possible, namely GR, positively curved spatial sections and a scalar field $\varphi$ in a self-interaction potential, we investigated the set of initial conditions under which a bounce might occur. For the bounce to take place, we assumed a potential of the form (\ref{V2}), with a constant term $V_0$, a symmetry-breaking mass parameter $\mu$ and a self-interaction coupling constant $\lambda$ and without loss of generality in this work, $\Lambda=0$.  This potential belongs to the more general, low-energy, phenomenological class of potentials in which all powers of the field could be present, including, \eg cubic terms. Such asymmetric terms are however not privileged, and we have found that in any case the main results of this paper are roughly independent of those extra terms. This, we interpret, is mostly due to the fact that all the physics of the bounce as well as that of the slow-roll contraction and expansion phases occurs near the top of the potential, the shape of which is hardly dependent on anything but the parameters we have used. The background we looked at here is therefore, as far as we can tell, generic enough that one be in a position to draw general conclusions on the evolution of the perturbations through the bounce, which was the main concern of the second part of this work.

Our first definite conclusion concerns the background and is provided by the phase space analysis.  It is found that, given positively curved spatial sections, the potential $V\left(\varphi\right)$ as given by Eq.~(\ref{V2}), and assuming the occurence of an inflationary phase some time in the history of the universe, a bounce {\it can} indeed have occured in the past with nonzero probability, preceded by a slow-roll contracting phase.  Note that the form of the potential we have chosen is the simplest one in which a bounce can happen and with which $\varphi$ stays bounded from below.  The flipside of this result is that given ``natural'' initial conditions in the far past, \ie a large but contracting universe with a scalar field slightly disturbed away from the minimum of its potential, trajectories for the evolution are highly unstable and fine-tuning is a necessity if one is to head towards a bounce and avoid singular behaviour.  The shape of the potential must also be ajusted, not unlike what is required in inflation, to ensure an inflationary phase lasting at least 60--70 e-folds.  At this stage, one should also mention that a slowly contracting precooling phase, during which, say, a radiation dominated universe moves towards a scalar field dominated universe, by some kind of condensation process, is a necessary step that should occur prior to the bouncing mechanism that we have described in this work.  The question of whether, after such a precooling phase, the scalar field dominated $\mathcal{K}=1$ universe will follow a bouncing rather than a singular trajectory is left open, not only in view of the first remark made in this paragraph, but also because one can choose to constrain the likelihood of such nonsingular trajectories by looking for their signatures in cosmological data (CMB and large structure formation for instance).  This naturally leads to the study of the perturbations and attempts to draw generic conclusions on their behaviour across the bounce for a given class of models.

Curvature perturbations, encoded in $\zeta$ (or $\zeta_{_\mathrm{BST}}$ in the $\mathcal{K}=1$ case) are often advocated to be conserved, \ie constant, in several of the known cosmological phases, such that they can be used as reliable estimators. Such a proposal frequently meets with strong disagreement.  In Ref.~\cite{Martin:2003bp} for instance, it was shown that out of the three reasons why  $\zeta_{_\mathrm{BST}}$ is usually conserved, two at least are not fulfilled, and the third is presumably meaningless in the cosmological bounce context. In the case at hand, and working in the observable wavelength (large) range, we found that $\zeta_{_\mathrm{BST}}$ is indeed not conserved, with both its spectrum and amplitude modified.  We used a matching procedure connecting a Bessel solution for the evolution of the variable $u$ prior to the bounce to another after the occurence of the bounce.  Given an initial quasiscale-invariant spectrum, the transfer function for the perturbations preserves the initial power-law to within order of the slow-roll parameters, but adds a leading order oscillatory term.  This result was confirmed by a full numerical evolution of the dynamical equation for $u$.  A priori, the form (\ref{cos2}) might lead to the hasty conclusion that this class of models can safely be excluded by CMB observations. We found however that in a large part of the parameter space in which a bounce is realized (see Section~\ref{sec:phase}), the spectrum of CMB multipoles (Fig.~\ref{osc}) is merely modified in a way reminiscent of superimposed oscillations, a possibility that has attracted some attention in recent years~\cite{MR07}, and was found to be not inconsistent with the data.  Any oscillations in the matter power spectrum or CMB multipoles are expected to be mostly smoothed out either by transfer functions or observational windowing functions. One is then led to predict, as in the completely different trans-Planckian case~\cite{Martin:2003sg}, some amount of superimposed oscillations which is not any worse (and might in fact be statistically better) than the power-law only inflationary prediction.   Constraining the parameters of the whole class of effective 4D bouncing models in which the potential term for the perturbations at the bounce has a negligible effect, using large-scale galaxy surveys and CMB data, remains to be done~\cite{LPRprep}. Once this is achieved, a complete model will need to be constructed, starting with an initial large contracting phase, with appropriate (physically justified) initial conditions, filled with matter and radiation, and evolving in such a way as to connect to the present model via a precooling phase in the recent $K-$bounce proposal~\cite{AP07}.  Such a mechanism is presently under investigation~\cite{precool:2008}.

\section*{Acknowledgments} 

We wish to thank R.~Abramo J.~Martin and N.~Pinto-Neto for many illuminating discussions.  We especially thank C.~Ringeval for providing the expected spectrum expressed in terms of temperature fluctuations.
Finally, FTF would like to thank the French/Brazilian cooperation CAPES/COFECUB for financial support.
\bibliography{references}
\end{document}